\documentclass[letterpaper,tight,classic,final,pageclassic,orcish,nobm]{iopart}
\usepackage{graphicx,graphicx,epsfig,amsfonts,amssymb} 

\begin{document}

   
\newcommand{\rr}{{\bf r}}
\newcommand{\uu}{{\bf u}}
\newcommand{\kk}{{\bf k}} 
\newcommand{\RR}{{\bf R}}
\newcommand{\qq}{{\bf q}}
\newcommand{\QQ}{{\bf Q}}
\newcommand{\GG}{{\bf G}}
\newcommand{\dd}{{\vec{\bf \delta}}}
\newcommand{\cc}{{\hat c}}
\newcommand{\ah}{{\hat a}}
\newcommand{\bb}{{\hat b}}
\newcommand{\znpk}{z_{n'k}}
\newcommand{\znk}{z_{nk}}
\newcommand{\ccd}{{\hat c^\dagger}}
\newcommand{\ahd}{{\hat a^\dagger}}
\newcommand{\bbd}{{\hat b^\dagger}}
\newcommand{\la}{\langle}
\newcommand{\ra}{\rangle}
\newcommand{\up}{\uparrow}
\newcommand{\dn}{\downarrow}
\newcommand{\rar}{\rightarrow}
\def  \bsig    {\mbox{\boldmath$\sigma$}}
\def  \btau    {\mbox{\boldmath$\tau$}}

\review{Stochastic pump effect and geometric phases in dissipative and stochastic systems}
\author{N.A. Sinitsyn}

\address{
  Center for Nonlinear Studies and 
  Computer, Computational and Statistical Sciences Division, 
  Los Alamos National Laboratory, Los Alamos, NM 87545 USA
}



\begin{abstract}{
The success of Berry phases in quantum mechanics stimulated the study of  similar phenomena in other areas of physics, including the theory of living 
cell locomotion and motion of patterns in nonlinear media. More recently, geometric phases have been applied to
systems operating in a strongly stochastic environment, such as molecular motors. We discuss 
 such geometric effects in purely classical dissipative stochastic systems and their role in the theory of the stochastic pump effect (SPE).
 }
\end{abstract}
\tableofcontents  

\maketitle
 
\section{Introduction}
The discovery of the  Berry phase \cite{berry-84} revolutionized the study of many quantum mechanical phenomena.
To explain this discovery, consider a quantum system with a Hamiltonian $\hat{H}({\bf k})$, where ${\bf k}={\bf k}(t)$ represents a vector of {\it control parameters} that 
can change with time along a prescribed path. The state of the quantum system is described by a complex valued wave function $\Psi$. However, 
the physical state itself only determines its wave function up to a phase because the wave functions $\Psi$ and $  e^{i\phi} \Psi$
 define the same physical state. We assume that, initially, the 
wave function is in one of the nondegenerate
eigenstates of the Hamiltonian $\hat{H}({\bf k})$.

\begin{figure}[h]
 \centerline{\includegraphics[width=1.2in]{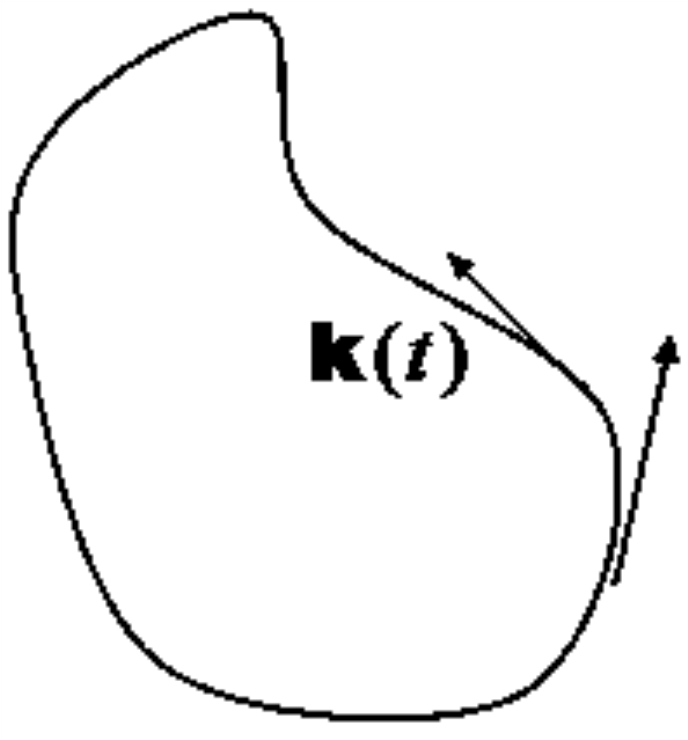}} 
  \caption{\label{contour} A closed contour in a control parameter space.} 
\end{figure}

 If the control parameters ${\bf k}$ change with time adiabatically slowly, then the Hamiltonian is explicitly time dependent 
but the {\it adiabatic theorem} of quantum mechanics \cite{berry-book} guarantees that the wave function will remain an instantaneous eigenstate of the Hamiltonian  $\hat{H}({\bf k}(t))$ during the evolution. 
Assume also that control parameters are changing along a closed countour in the parameter space,
so that at the end of the evolution they return to the initial values,
as in
Fig.~\ref{contour}. 
According to the adiabatic theorem, after completing the cycle, the physical state of the system should 
coincide with the initial one.
 This theorem,  does not mean that the phase of the wave function returns to the initial value. Careful examination shows \cite{berry-84} that the phase picked up after a cyclic evolution can be written as a sum of two components
\begin{equation}
\phi = \phi_{\rm dyn} + \phi_{\rm B},
\label{gaugepsi2}
\end{equation}
 where the  dynamic phase $\phi_{\rm dyn} =- \int_0^TE(t)dt$  appears even when parameters are fixed, $T$ is time of cyclic evolution, $E(t)$ is energy detemined as the
 instantaneous eigenvalue of the Hamiltonian $\hat{H}({\bf k}(t))$, and we assume that $\hbar=1$. The second
 contribution in (\ref{gaugepsi2}) is called the {\it Berry phase}. It has no stationary counterpart and is purely geometrical, in the sense that it depends only on the choice of the Hamiltonian and on the path in the parameter space.
 In particular, it
 does not depend explicitly on  time of the evolution $T$ or on the rate of motion along the contour.  Given the dependence of the 
eigenstate $\vert u \ra$ of the Hamiltonian
 on ${\bf k}$ in some gauge, the Berry phase is given by
\begin{equation}
\phi_{\rm B} = \oint_{{\bf c}} {\bf A} \cdot {\bf k},\quad {\bf A} = \la u \vert i \partial_{{\bf k}} u \ra, 
\label{berry-phase} 
\end{equation}
where ${\bf c}$ is the contour in the space of control parameters, and ${\bf A}$ is called {\it the Berry connection} (see  David J. Griffiths \cite{grifits-book} for a pedagogical derivation). 
The cyclic Berry phase is gauge invariant and has measurable effects, which have been confirmed experimentally. 

Berry's discovery has a rich prehistory. 
As Berry himself pointed out \cite{berry-anticipation}, a number of effects in quantum mechanics had been related to the unusual phase evolution of the wave function even before the Berry's discovery \cite{berry-84}.
Other historical antecedents can be found, for example, in the first efforts to create a semiclassical theory of the anomalous Hall effect \cite{smit-1,smit-2}. 
However, it was Berry's work that unified all these known geometric effects under a universal theoretical framework. This framework has proved very useful in uncovering new  effects and 
has provided the theoretical groundwork for entire branches of physics such as topological quantum computations \cite{tregubovich}, the quantum theory of polarization \cite{resta} and
 the theories behind various extraordinary Hall effects \cite{sinitsyn-rev,murakami-she,sinova-she}.

The Berry phase is an example of the {\it anholonomy} effect encountered in the theory of differential equations.
 Anholonomy can be non-rigorously defined as failure of vectors return to their initial values after a parallel transport along a closed contour.

\begin{figure}[h]
 \centerline{\includegraphics[width=2.4in]{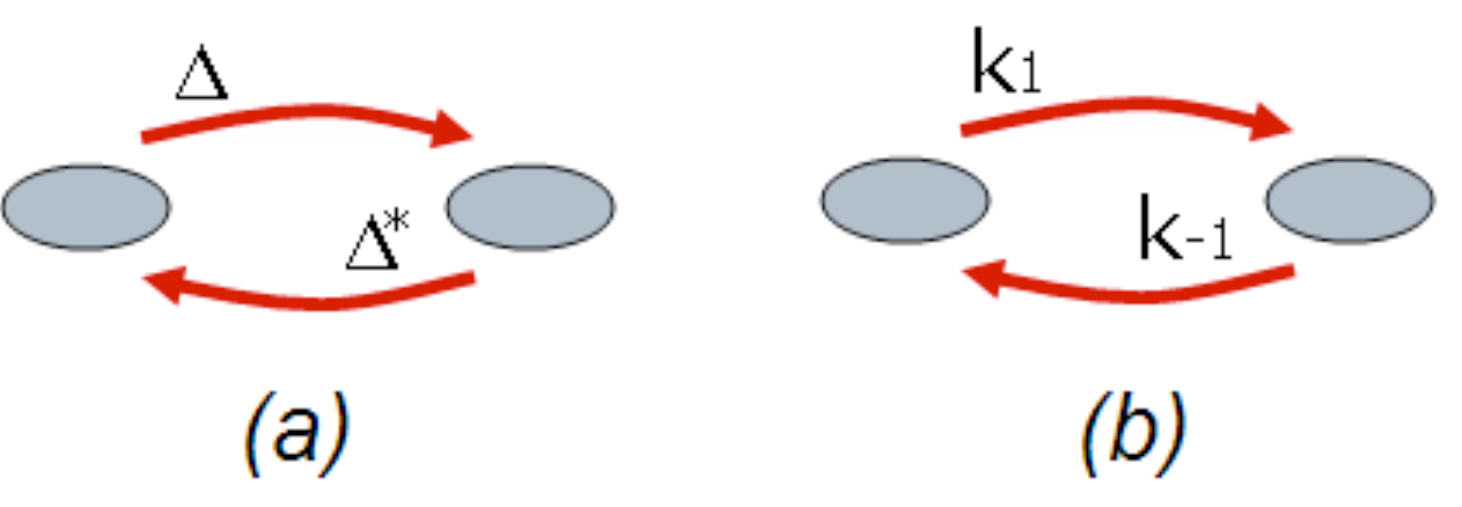}} 
  \caption{\label{two-state} Two-state systems: (a) quantum (b) stochastic.}  
\end{figure}

Since there are both quantum mechanical and classical mechanical antecedents of the geometric phase, it is natural to ask whether there are also   
 antecedents in stochastic processes. Here, we should point out that there is a fundamental difference from quantum mechanics,
 which prevents  direct analogies. While  the state of a quantum system defines the wave function only up to an overall phase, 
a classical ergodic stochastic system can  be described by a probability vector, without allowing any additional freedom in its definition.
 For example,
consider a {\it Markov process}, i.e a stochastic process whose future evolution depends only on the present state but does not depend on the past.
 The evolution of the probability vector satisfies equations which are reminiscent of the quantum mechanical evolution.
 Fig.~\ref{two-state} shows the simple example of a  2-state quantum system and its stochastic counterpart. In both cases the evolution is described 
by linear differential equations with  $2\times 2$ evolution matrices. In the quantum mechanical case, the amplitudes $u_1$ and $u_2$ of two states evolve according to the Schr\"odinger equation
\begin{equation}
i\frac{d}{dt}\left( \begin{array}{l} 
u_1 \\
u_2
\end{array} \right) = \left( \begin{array}{ll}
\epsilon_1(t) & \Delta(t) \\
\Delta^*(t) & \epsilon_2(t) \end{array} \right) \left( \begin{array}{l}
u_1 \\
u_2 
\end{array} \right),
\label{shrod}
\end{equation} 
while the continuous 2-state Markov process, with probabilities $p_1$ and $p_2$ of the first and the second states, is defined by differential system \cite{lax}
\begin{equation}
\frac{d}{dt}\left( \begin{array}{l} 
p_1 \\
p_2
\end{array} \right) = \left( \begin{array}{ll}
-k_1(t) & \, k_{-1}(t) \\
\,\, k_1(t) & -k_{-1}(t) \end{array} \right) \left( \begin{array}{l}
p_1 \\
p_2
\end{array} \right),
\label{shrod-stoch}
\end{equation} 
where parameters $k_1$ and $k_{-1}$ are called {\it kinetic rates}. They describe how often a system jumps from one state into the other. 
In spite of the similarity of Eqs. (\ref{shrod}) and (\ref{shrod-stoch}),
it is well known that a quantum mechanical, cyclically driven 2-state system
can have a nontrivial geometric Berry phase, but that is not true for its stochastic counterpart. For given values of parameters, an ergodic Markov chain has a unique steady state.
Under slow evolution of parameters, the probability vector will simply follow the path of instantaneous steady state values, returning to the
initial vector after the parameters complete a full cycle. 

\begin{figure}[h]
 \centerline{\includegraphics[width=1.8in]{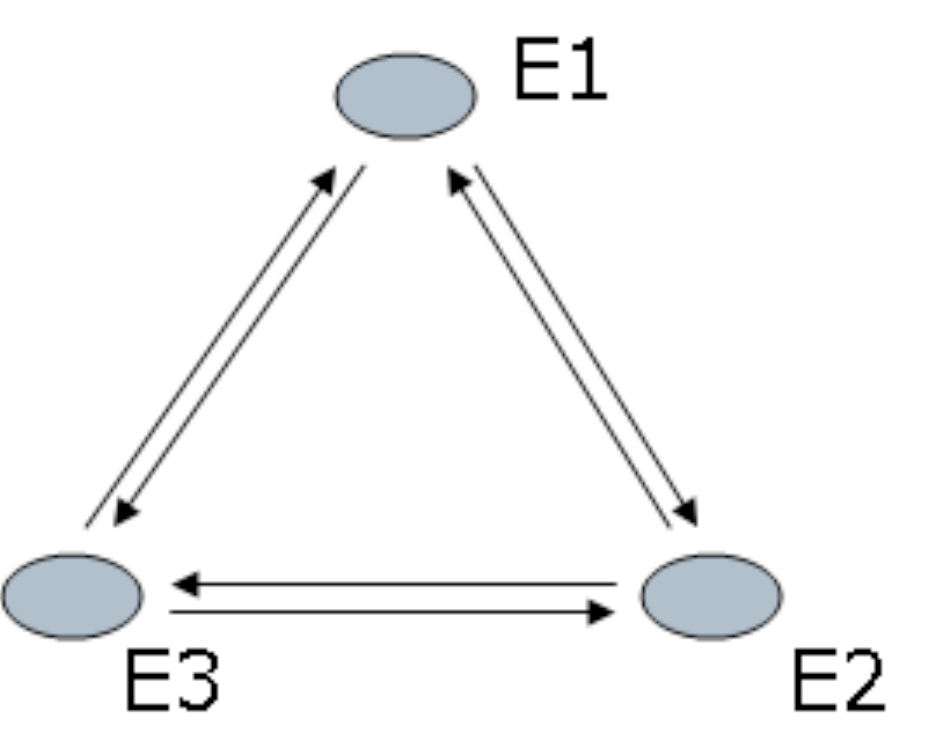}} 
  \caption{\label{three-state} A 3-state Markov chain. This model can be considered as a minimal model to describe the stochastic behavior of a molecular motor in Fig.~\ref{f0f1}.} 
\end{figure}

This example shows that, in an obvious sense, the Markov process does not lead to adiabatic geometric phases. However, this conclusion is restricted only to the evolution of the  probability vector for a present state. There are other
characteristics describing stochastic processes. For example, one can consider stochastic transitions among 3 states in Fig.~\ref{three-state} and ask what is the probability that the system makes exactly $n$ full cycles in a clockwise direction by the
given time $t$. In the following sections, we will derive the evolution equations for similar quantities and show that geometric phases do play an important role in their evolution.

Anholonomies also play a central role in classical thermodynamics. The Carnot cycle is an example of a thermodynamic process exhibiting anholonomy \cite{hannay-carnot}. 
Statistical properties of a
 system in thermodynamic equilibrium can be specified by a set of parameters, such as the volume,
 the pressure and the temperature. Slow cyclic changes of these parameters merely produce cyclic changes of the equilibrium properties, so, as in the example of a 2-state Markov chain, a driven
system returns to the initial state in a statistical sense by the end of a cycle. However, if one looks at the same process from a more general point of view, namely including effects of
 this process not only on the given system, but also 
on systems in contact with it,
a cyclic adiabatic evolution  of control parameters usually does not lead to the same finite state in the full phase space. The laws of  thermodynamics predict that the system  
converts part of the absorbed energy into production of the work.
 Moreover, for adiabatically slow evolution, the work produced depends only on the choice of the contour in the  parameter space,
but depends neither on
the rate of motion nor on
the mechanism of coupling to the environment. This means that the work can be expressed as a contour integral over the path in the space of control parameters. 
For example, for a gas in a reservoir with variable volume $V$ and temperature $T$, the work  $W$ produced per cycle is 
\begin{equation}
W = \oint_{{\bf c}(T,V)} p(T,V) \cdot dV,
\label{work} 
\end{equation}
where $p(T,V)$ is the pressure.

The property of dynamic systems to change their state in response to a periodic perturbation with zero bias has 
been widely used  in control theory \cite{control-1}. Look e.g. at the simple
input/output model described by differential equations
\begin{equation}
\dot{x}^i=f^i_j({\bf x}) u_j(t), 
\label{control1}
\end{equation} 
where $f_j$ are smooth functions of ${\bf x}$, and $u_j(t)$ represent control parameters. One can consider a simple periodic evolution in the control parameter space by setting 
 $u_i = \delta_{i,1}$ during an
infinitesimal time interval $t$, where $\delta_{i,j}$ is $1$ when $i=j$ and $0$ otherwise. After this, one sets $u_i=\delta_{i,2}$ during the following time interval $(t,2t)$, then takes
 $u_i=-\delta_{i,1}$ for $(2t,3t)$, and finishes with $u_i=-\delta_{i,2}$ during
$(3t,4t)$.
\begin{figure}[h]
 \centerline{\includegraphics[width=2.in]{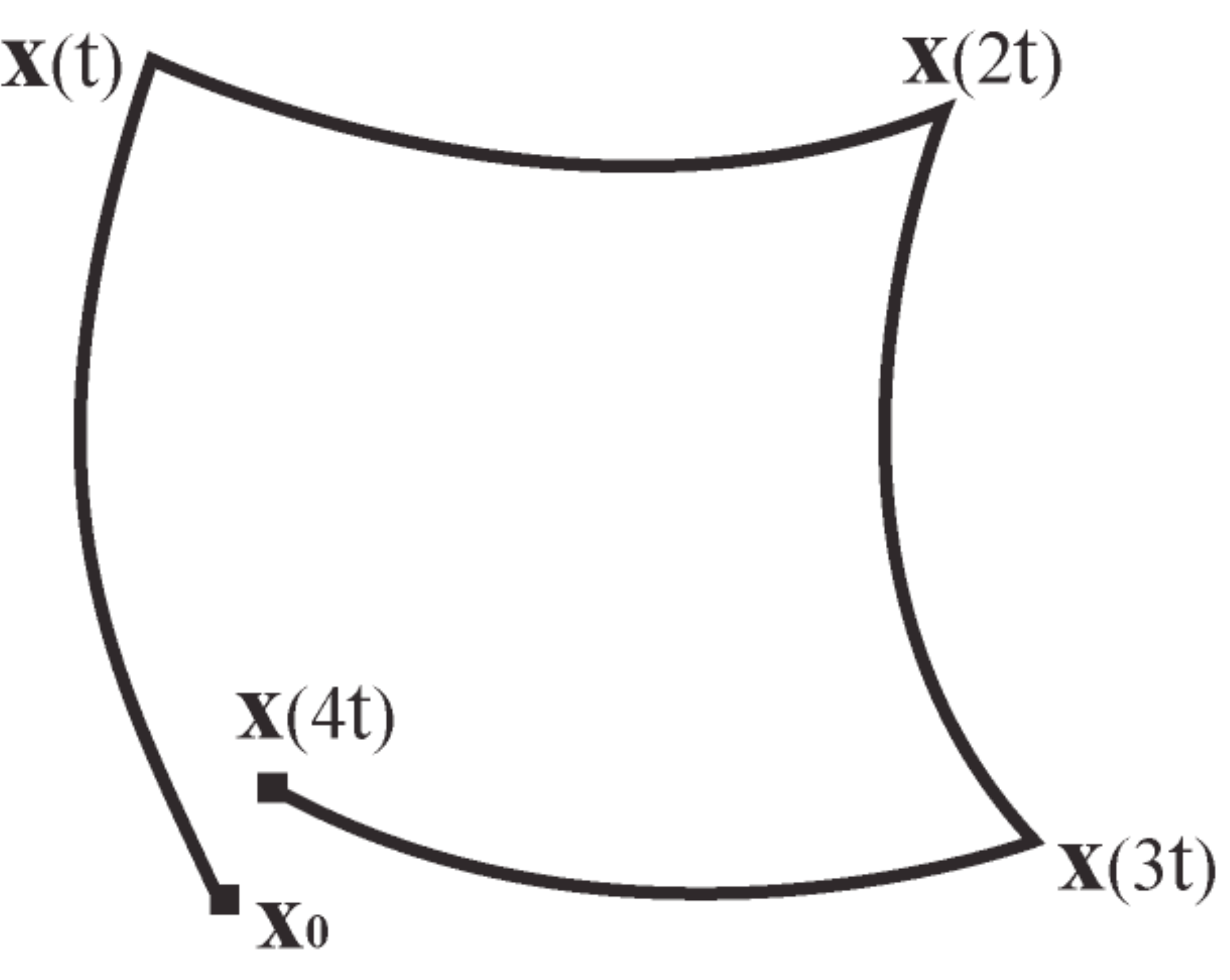}} 
  \caption{\label{pathfig} Trajectory of  ${\bf  x}$ in response to a periodic infinitesimal path in the space of control parameters $\{ u_i \}$.} 
\end{figure}


An easy perturbative calculation shows that, after one such an infinitesimal  evolution in the space of control parameters with zero bias, the vector ${\bf x}$ does not generally return to the initial state but rather acquires an additional correction, namely 
\begin{equation}
{\bf x}(4t) ={\bf x}_{0}+t^2 \left[ {\bf f}_1,{\bf f}_2 \right]( {\bf x}_0) +{\rm O}(t^3),  
\label{corr1}
\end{equation}  
where the operation
$[{\bf f}_1,{\bf f}_2]^{i}=(f_1^{j} \partial f_2^{i}/\partial x^{j})- (f_2^j \partial f_1^i /\partial x^j)$ is called the {\it Lie brackets} of the vector fields ${\bf f}_1$ and ${\bf f}_2$.
In general, the the
value at ${\bf x}_0$ of the Lie bracket $[{\bf f}_1,{\bf f}_2]$ can even be
linearly independent of ${\bf f}_1({\bf x}_0)$ and ${\bf f}_2({\bf x}_0)$. This phenomenon is
exploited in Chow's theorem which, together with the theorem of
Frobenius,
describes the space of configurations that can be reached using only a
prescribed set of vector fields to get around 
 \cite{robot-book}.

Although the history of anholonomy effects reaches at least as far back as the 19th century, a complete survey of its development is beyond the scope of this article and we will not pursue it further. 
Instead, we will concentrate on applications of geometric phases to stochastic and dissipative processes. These topics have attracted  attention
relatively recently due, in part, to the success of the Berry phase in quantum mechanics.   We attempted to make the review accessible to an audience not familiar with 
the mathematical
 theory of fiber bundles.
This review is also not about the large body of work related to decoherence effects on quantum mechanical Berry phases or on
geometric phases in chaotic non-dissipative systems.
The reader should 
consult  \cite{berry-book,jarzynski-chaos,kuvshinov,gefen,niu-book} for an introduction to these topics.

The structure of this review is as follows.
 Section 2 reviews  extensions of the Berry phase idea to non-unitary evolution.
Section 3 describes the theory and applications of geometric phases in dissipative systems with a continuous symmetry of steady state solutions. One important application is to
control of pattern motion in nonlinear media. Section 4 briefly reviews the geometric theory of locomotion of micro-organisms. 
In section 5, we introduce the stochastic pump effect and its relation to  geometric phases in the evolution of the 
moment generating functions. In section 6, we generalize the concept of the geometric phase in stochastic processes
to  noncyclic evolutions. In section 7, we explain how geometric phases can influence the kinetics of slow variables in a coarse-grained
description of a stochastic process with a hierarchy
 of important time scales.
This is regarded as a
stochastic analog of the quantum mechanical Born-Oppenheimer
approximation. We also
study stochastic analogs of Berry phases from this point of view.
The reader who is not interested in complicated mathematical details might prefer to skip  section 7 on a first reading.
In section 8, we review the geometric phases in the ``limit cycle'' evolution.
Section 9 is about constraints that detailed balance conditions  impose on geometric effects in systems near thermodynamic equilibrium.
In section 10,  we apply some of the techniques we have discussed to the theory of molecular motor operations. In section 11, we outline several  directions for possible future research.

\section{Geometric phases in non-unitary evolution}
Non-unitary evolution is often considered in quantum mechanical problems, where the coupling to the environment
 is described phenomenologically by introducing extra parameters in the equations of motion for a density matrix or a wave function.
In many physical problems one can encounter evolution equations, similar to quantum mechanical ones but with a non-Hermitian operator replacing the Hamiltonian. Examples  can be found in electronic circuits \cite{stehmann}, 
optics \cite{berry-dennis}, acoustics \cite{shuvalov}. The evolution of any dissipative system near a stable point or a limit cycle can be linearized and assume a form similar to the quantum mechanical Schr\"odinger equation
for a state vector.
Motivated by the success of the quantum mechanical Berry phase, several studies \cite{garrison-88,nonherm-bf1,berry-nonherm,garanin} were devoted to  geometric phases in 
systems with non-Hermitian Hamiltonians. 
The models considered could generally be written in the form 
\begin{equation}
\frac{d}{dt} \vert  u \ra = \hat{H}({\bf k}) \vert  u \ra,
\label{nonherm1}
\end{equation}  
where $\hat{H}$ is  arbitrary $N \times N$  matrix,
 and $\vert  u \ra$ is an $N$-vector.  

Imitating the derivation of the quantum mechanical Berry phase in the adiabatic limit leads one to a similar result: 
if the vector $\vert u \ra $ at the initial moment of the evolution is one of the eigenstates of the matrix $\hat{ H}({\bf k})$, i.e. if 
\begin{equation}
\hat{H} \vert u(0) \ra = \varepsilon \vert u(0) \ra,
\label{nonherm2}
\end{equation} 
then after a slow cyclic evolution of parameters, the vector $\vert u \ra$ generally returns to the initial one up to a factor
\begin{equation}
\vert u (T)\ra= e^{-\oint_{{\bf c}}{\bf A}\cdot d{\bf k}} e^{\int_0^T dt \varepsilon (t)} \vert u(0) \ra,
\label{nonherm3}
\end{equation}
which can be separated into  dynamic and geometric parts.
Strictly speaking, the values in the exponents in (\ref{nonherm3}) are not phases because they are no longer purely imaginary. However, it is generally accepted to use the  term, ``phase'', because of the strong analogy with 
quantum mechanical phases. 
Berry \cite{berry-nonherm} pointed that it is convenient to express the geometric phase by introducing also the left-eigenstates of the Hamiltonian $\hat{ H}({\bf k})$, such that
\begin{equation}
\la u \vert \hat{H} =  \la u \vert  \varepsilon.
\label{nonherm4}
\end{equation}
The eigenvalues for left and right eigenvectors coincide.  
For a non-Hermitian Hamiltonian, the eigenvalues are no longer
real and the components
 of the left eigenvector are no longer complex conjugates of the right eigenvector. In other respects, there is a strong similarity with quantum mechanical Berry phases, e.g.
the connection ${\bf A}$ can be written as
\begin{equation}
{\bf A} =\frac{\la u \vert \partial_{{\bf k}}  u \ra }{\la u \vert u \ra}.
\label{nonherm5}
\end{equation}     

Non-unitary evolution does not necessarily describes a dissipation. The geometric phases with a non-Hermitian 
Hamiltonian that realizes transformations of the group $SU(1,1)$  (i.e. transformations that preserve the form $|z_1|^2-|z_2^2|$ of
a complex valued 2-vector $(z_1,z_2)$), have attracted considerable attention because of their applications to squeezed states \cite{sl2r-1,sl2r-2,sl2r-3,sl2r-4}, and a number of 
models in classical mechanics and optics  \cite{han-99,han-89,hannay-85}.  
The group $SU(1,1)$ is a covering group of the 3-D Lorentz group. The corresponding geometric phase is responsible for a variety of relativistic effects, such as 
Thomas precession \cite{littlejohn-88,ferraro-00}. We refer the interested reader to the  review \cite{mukunda-03jpa}.
More important for our subject is that the group $SU(1,1)$  is isomorphic to the group $SL(2,R)$ of $2\times 2$ matrices with real entries and the unit determinant. The $SL(2,R)$
 evolution  can be used to describe a classical dissipative
 system. 
Corresponding geometric phases were studied both theoretically and experimentally in connection with  light propagation through a set of  polarizers \cite{sl2r-optics1,sl2r-optics2,sl2r-optics3}.
 Recently, the relation of this geometric phase
to the stochastic pump effect was discussed in \cite{sinitsyn-08jpa2}.

The geometric phase in Eq. (\ref{nonherm3}) was generalized to  nonabelian and non-adiabatic evolutions \cite{garanin,nonherm-nonadiab1} and a number of applications were proposed,
e.g. to optically active refracting media \cite{berry-nonherm}. 
One interesting property of non-Hermitian Hamiltonians, not found in quantum mechanics, is the possibility of so-called exceptional points in the spectrum.
If a contour encloses such points, the eigenvalues of the Hamiltonian
 are not single valued along this contour. A simple example is the following $2\times2$ matrix, which depends on a complex parameter $z$ and which has $0$ as an exceptional point:
$$
\left(
\begin{array}{ll}
0&1 \\
z&0
\end{array}
\right), \quad z \in \mathbb{C}-\{ 0 \}.
$$
Its eigenvalues $\lambda_{\pm}=\pm\sqrt{z}$ are double valued functions on the space $\mathbb{R}^2 - \{ 0 \}$.
Encircling such points, the eigenvectors acquire a geometric phase of a new type which has been studied theoretically in  
\cite{mostafazadeh,mailybaev,gunter, sun} and observed in experiments \cite{dembrovski-04,dembrovski-01}. 

In spite of this progress,
one might think that the geometric phase in dissipative evolution would be insignificant in comparison with the dynamic part in (\ref{nonherm3}) on the grounds that the latter
 becomes either exponentially large or exponentially small with time. However, not all modes in dissipative evolution grow or decay exponentially since  
the matrix $\hat{H}$ might also have zero modes, i.e. one or several linearly independent states with a zero eigenvalue. If all other modes decay quickly, according
to (\ref{nonherm3}), the evolution of such a dissipative system in the adiabatic limit should instead be governed by the geometric phases. More generally, geometric phases can also be important 
when, for fixed values of parameters, a system relaxes to a limit cycle. In that case, the corresponding eigenvalue is purely imaginary.

\section{Control over pattern position and orientation}
In early 1990s, Landsberg \cite{landsberg-92, landsberg-93}, and independently Ning and Haken \cite{haken-1,haken-2}, suggested that geometric phases should generally appear in many classical 
 systems which can be described by a set of nonlinear differential equations with a one-parameter group of symmetry transformations. In such a system, it is possible to reduce the evolution equations to a form
in which one of the variables does not affect the evolution of the others, that is
\begin{equation}
\frac{d{\bf Y}}{dt}={\bf F}({\bf Y},{\bf \lambda}),\quad \frac{d\Theta}{dt}=H({\bf Y}),
\label{eq_lan1}
\end{equation}  
where $ \Theta$ and the vector ${\bf Y}$ represent generalized coordinates of the system while ${\bf F}$ and $H$ are nonlinear functions of the coordinate vector ${\bf Y}$, but not of $\Theta$.
Assume that the system is initially at the steady state and that the parameter vector ${\bf \lambda}$ changes slowly with time. Landsberg considered the case when the evolution of the variable ${\bf Y}$ is
 dissipative, so that if parameters ${\bf \lambda}$ are time-independent, ${\bf Y}$
relaxes to a steady state value ${\bf Y}^*({\bf \lambda})$. In that case, for adiabatically slow evolution of ${\bf \lambda}$,
 ${\bf Y}$ simply follows a quasi-steady state trajectory up to a small non-adiabatic correction  
\begin{equation}
{\bf Y}(t) \approx {\bf Y}^* + D{\bf F}^{-1}({\bf Y}^*)\partial_t{\bf Y}^*, 
\label{follow}
\end{equation}
where  $D{\bf F}=\left(\partial {\bf F}/\partial {\bf Y} \right)_{{\bf Y}={\bf Y}^*}$ is the linearization of the vector function ${\bf F}$,
and $D{\bf F}^{-1}$ is the inverse of the linear function $D{\bf F}$.
 However, one should not assume that $\Theta$ can be uniquely determined by given values of the control parameters. Due to the symmetry  $\Theta \rightarrow \Theta+\delta \theta$ of Eq. (\ref{eq_lan1}), where $\delta \theta$ is
arbitrary constant,
 the steady state value of $\Theta$ is not specified. Hence $\Theta$ does not have to return to its initial value after the  parameters ${\bf \lambda}$ complete a full cycle.
Landsberg showed in \cite{landsberg-93} that, after a cyclic evolution, the variable $\Theta$ changes by an amount given by a trivial dynamic part $\Delta \Theta_{\rm dyn}=\int dt H({\bf Y}^*(t))$ plus a geometric  contribution
\begin{equation}
\Delta \Theta_{\rm geom} = \oint_{{\bf c}} {\bf A}\cdot d{\bf \lambda},\quad {\bf A}= DH({\bf Y}^*)D{\bf F}^{-1}({\bf Y}^*) \partial_{\bf \lambda} {\bf Y}^*, 
\label{eq_lan2}
\end{equation}
where $DH=\partial H/\partial{\bf Y}\vert_{{\bf Y}={\bf Y}^*}$ is the linearization of the function $H$ near the point ${\bf Y}^*$. 

Suppose, that instead of a vector ${\bf Y}$ with discrete set of entries, we deal with continuous systems. The vector index then becomes a continuous coordinate and ${\bf Y}$ is a function of this coordinate.
To generalize (\ref{eq_lan2}) to such continuous systems, Landsberg considered equations of the form
\begin{equation}
\frac{d\Psi(t,x)}{dt}= \hat{F}(x, {\bf \lambda})\Psi(t,x), 
\label{eq_lan3}
\end{equation}  
where $\hat{F}$ is now a nonlinear operator, which depends on time only through time-dependent control parameters ${\bf \lambda}$, and where the  evolution equation (\ref{eq_lan3}) is assumed to be invariant 
under a continuous group $G$ of symmetries, such as the group of  translations in the direction of the coordinate $x$. 

Let $\vert \psi(x) \ra$ be a {\it stationary pattern profile}, i.e. a time-independent solution of (\ref{eq_lan3}) for constant ${\bf \lambda}$.
Let $D\hat{ F}$ be the differential operator, which is the linearization of the operator $\hat{ F}$ near the solution $\vert \psi(x) \ra$, and  let $\langle v_0|$
be the zero mode of the  differential operator $D\hat{ F}^{+}$, which is the
{\it
conjugated differential operator} to $D\hat{ F}$, in the sense that $\int g(Df)=\int (D^+g)f$. 
Landsberg showed that after a cyclic evolution in the parameter space, the stationary pattern will change by a geometric shift $\Delta\Theta_{\rm geom}$, given by
\begin{equation}
\Delta \Theta_{\rm geom} = \oint_{{\bf c}} {\bf A}\cdot d{\bf \lambda},\quad {\bf A}= -\frac{\langle v_0|\partial_{{\bf \lambda}} \psi \rangle}{\langle v_0|\hat{\chi} \psi\rangle },
\label{eq_lan4}
\end{equation}
along the symmetry direction, where $\hat{\chi}$ is the generator of transformations of the symmetry group $G$. 
\begin{figure}[h]
 \centerline{\includegraphics[width=2.0in]{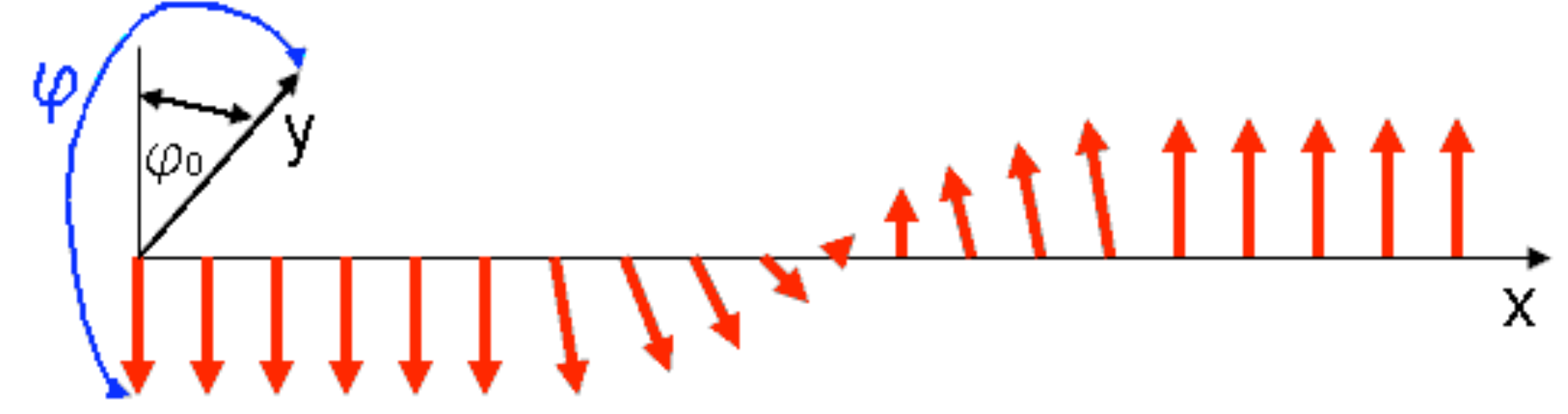}} 
  \caption{\label{DW} The wire with a hard anisotropy axis along it and a domain wall between two opposing magnetization directions.
 Easy axis anisotropy is perpendicular to the $x$-direction, and makes an angle $\varphi_0$ with a $y$-axis. Red vectors show the direction of the local magnetization.} 
\end{figure}

Eq. (\ref{eq_lan4}) can be illustrated by a following simple example  \cite{coullet-91,sinitsyn-08prb}. 
Consider a 1D ferromagnetic wire with a strong hard axis along it, which favors a magnetization direction transverse to the wire, as shown in Fig.~\ref{DW}.
Assume that a fixed set of rectangular coordinate axes has been chosen, with the $x$-axis along the wire. 
 We also assume the presence of a
 weak transverse anisotropy, so that  the magnetization energy is described by the energy functional
\begin{equation}
E \approx \int dx \{ J(\partial \varphi/\partial x)^2 + K \sin^2(\varphi-\varphi_0) \},\\
\label{DF-2}
\end{equation}
where $\varphi$ is the magnetization angle with the $y$-axis, and 
 the parameter $\varphi_0$ is the angle that the transverse anisotropy axis makes with $y$-axis (Fig.~\ref{DW}).

In the limit of strong dissipation,  the evolution of the variable $\varphi(x,t)$ is given by 
\begin{equation}
\alpha \partial_t\varphi =-\frac{\delta E}{\delta \varphi}= 2J\partial_x^2 \varphi -K\sin[2(\varphi-\varphi_0)], %
\label{relax}
\end{equation}
where $\alpha$ is a damping constant.
Note that Eq. (\ref{relax}) is invariant under the translation $x\rightarrow x-\delta x$, which justifies the use of the geometric theory of \cite{landsberg-92}.
The generator of this symmetry is $\hat{\chi}=-\partial_x$.

At equilibrium, the ground state of (\ref{DF-2}) is doubly degenerate at $\varphi=\varphi_0$ and $\varphi=\varphi_0+\pi$.
Consider the stationary solution of (\ref{relax}) describing a {\it domain wall}  connecting these two states
\begin{equation}
\varphi^{\rm dw}(x;\varphi_0,x_0)=\varphi_0 + 2\tan^{-1}e^{(x-x_0)/\Delta}, \Delta=\sqrt{J/K_0},
\label{dw2}
\end{equation}
where $x_0$ and $\Delta$ can be called respectively the {\it position} and the {\it size} of the domain wall.
Assume that the parameter $\varphi_0$ is slowly time-dependent and varies from $0$ to $2\pi$ as the transverse anisotropy axis performs one rotation around the $x$-direction. 
One can realize this situation, for example, by physically rotating a wire.
In  our model,  $\varphi_0$ is the control parameter, and Eq. (\ref{eq_lan4}) gives
\begin{equation}
\delta x_0 = \oint \frac{ \int_{-\infty}^{\infty} dx[ v_0(x) \partial_{\varphi_0} \varphi^{\rm dw}(x)]}{ \int_{-\infty}^{\infty}dx[ v_0(x)\partial_x \varphi^{\rm dw}(x) ] } d\varphi_0,
\label{lands}
\end{equation}
where $v_0(x)$, given by 
\begin{equation}  
v_0(x)=\frac{\Delta}{\cosh \left( (x-x_0)/\Delta \right)},
\label{vo}
\end{equation}
is the zero mode of the self-adjoint operator
\begin{equation}
D\hat{F}=D\hat{F}^+=2J\partial_x^2-2K\cos[2(\varphi^{\rm dw}(x;\varphi_0,x_0)-\varphi_0)].
\end{equation}
Substituting this value of $v_0(x)$ into (\ref{lands}) we find
\begin{equation}
\delta x_0 = \int_0^{2\pi} A_{\varphi_0}d\varphi_0,\,\,\,\,\, A_{\varphi_0} = \pi \Delta/2,
\label{shift-2}
\end{equation}
which was derived in \cite{sinitsyn-08prb} using the secular perturbation theory.
 
A number of theoretical and experimental studies of the motion of domain walls in liquid crystals under the influence of periodic perturbations have been performed
 \cite{coullet-91,frisch-94,rudiger-07,kawagishi-95,vierheilig-97}, but the role of the geometric phase  has not been discussed. 
Landsberg's theory was aimed at control of wave patterns in nonlinear media. Such  control was discussed in more detail for specific applications to nonlinear chemical reactions 
\cite{rudzick-06}, nonlinear optics \cite{scroggie-05}, hydrodynamics \cite{abarzhi-07} and semiconductor microresonators \cite{maggipinto-00}. Optical applications proposed by Ning and Haken were extended
by Toronov and Derbov \cite{ toronov-1,toronov-2}.

Studies of special problems with a mathematical structure similar to the Landsberg-Ning-Haken formalism can be found  even prior to Refs.~\cite{landsberg-92,landsberg-93}.
For example, the use of a rotating electric field was suggested as a means of separating chiral molecules in solution  \cite{pomeau-1,pomeau-2}.
Recently this idea was extended to the gaseous state, but the proposed effect is expected to be observed
in the non-adiabatic regime \cite{spivak-08}.  Geometric phases can also contribute to the  anomalous shift of
the trajectory of a magnetic bubble
 in a rotating nonuniform magnetic field, which is
called the {\it skew-deflection effect}
 \cite{slonczewski-book}.

Control over the motion of domain walls and other topological defects has been extensively studied in magnetic materials
 \cite{slonczewski-book,clarke-08,tretjakov-08,thiele-73}. 
The example of the domain wall shown above demonstrates that the projection of dynamics on the collective degrees of freedom should be performed with extra care to account for possible geometric phase effects.
For example, in the micromagnetics
literature,
 one often finds that the equations of motion for the collective coordinates ${\bf \xi}$ read \cite{tretjakov-08}
\begin{equation}
-\partial U/\partial {\bf \xi}- \Gamma \dot{{\bf \xi}}+G \dot{{\bf \xi}}=0,
\label{micromagn}
\end{equation}
where $-\partial U/\partial {\bf \xi}$ is the generalized force, $\Gamma$ is the symmetric dissipation matrix and $G$ is the antisymmetric gyrotropic matrix. However, Eq. (\ref{micromagn}) can lead to problems.
In the case of a domain wall (\ref{dw2}),
one can attempt to work with a single 
collective coordinate representing the position of the domain wall, and to regard the angle
 $\varphi$ and the size of the domain wall $\Delta$ as fast variables.
This choice might seem natural in
view of the fact that translation is the only continuous symmetry in the model and should dominate the physics at low energies and under slow perturbations. 
However, in the model under consideration, the wall was moving  because some parameters became time-dependent.
In the static case, there are no forces on the chosen collective degrees of freedom. 
Thus Eq. (\ref{micromagn}) can acquire extra geometric terms in explicitly time
dependent situations, and is therefore not, in itself, sufficiently
general even when only slow changes of parameters are considered. 
As another word of caution, we note that 
the translational symmetries in real applications are only approximate for standard magnetic materials because of the presence of impurities and discreteness of the lattice. 
As discussed in \cite{sinitsyn-08prb}, this is a serious obstacle to 
  purely geometric control over magnetic defects in practical applications. 

\section{Self-propulsion at low Reynolds numbers}
One of the first applications of geometric phases in dissipative systems was  proposed  by Shapere and Wilczek \cite{wilczek-88,shapire-88} in their description of  locomotion of microscopic organisms in a viscous fluid.
This theory was based on the well known observation that the motion of living organisms at low Reynolds numbers is, in fact, geometrical. 
Microscopic living organisms propel themselves in a liquid by performing periodic changes of their shapes. These changes correspond to noncyclic motion in a larger space, whose points describe the shape of the body, its position and its
 overall orientation. 
One can choose a {\it gauge} in this phase space, i.e. a rule to determine the position and the orientation of the body of any given shape with respect to a fixed coordinate system in the 3-dimensional space. Using such a gauge 
one can characterize the state of the body 
by a set of coordinates $(x,\alpha)$, where $\alpha$ is a vector of
shape parameters and $x=({\bf r},\phi)$
consists of the position ${\bf r}$ and orientation $\phi$ of the body.

 Body shapes are assumed to be directly controllable by the organism  subject to certain constraints such as
 conservation of volume, which allow only finite and quasi-periodic changes of ${\bf \alpha}$. The body interacts with a high viscosity liquid which is described by the incompressible 
Navier-Stokes equation. This equation should be solved  with no-slip boundary conditions at the surface of the organism to guarantee the absence of  force and a torque on it. 
For slow changes of shape, these no-slip conditions are expressible in
terms of the first partial
derivatives of $x$ with respect to time. In particular, they are
automatically satisfied by non-moving bodies.
After eliminating the  degrees of freedom of the liquid by solving the Navier-Stokes equation,  the equation that connects changes of $x$ and ${\bf \alpha}$ follows from the boundary conditions and has the form
\begin{equation}
dx={\bf A} ({\bf \alpha}) \cdot d{\bf \alpha},
\label{loc1}
\end{equation}  
where the connection ${\bf A}$ is defined  on the space of body shapes. Integration over a  closed path ${\bf c}$ in the space of body shapes leads to the purely geometric 
result that $\delta x = \oint_{{\bf c}} {\bf A} ({\bf \alpha}) \cdot d{\bf \alpha}$.  

The geometric theory of locomotion at low Reynolds numbers has found too many applications to summarize here. Fortunately, 
fairly good introductions and reviews are already 
available \cite{purcell-77,lowRe-book,lowRe-review}. Here we only mention that 
the theory was applied to determine optimal protocols for cell body changes
\cite{avron-04, najafi-pre04,astumian-aip,avron-07,golestanian-08prl}. It was found that such optimal moves are similar of those observed in some organisms \cite{avron-04,avron-07}. 
Discussions of simple illustrative models can be found in \cite{najafi-pre04,dreyfus-05}. An application of the theory  for the propulsion of microscopic objects by manmade motors
 at low Reynolds numbers can be found in \cite{dreyfus-nat05}.

\section{Stochastic pump}

A {\it stochastic pump} is a stochastic system that responds with nonzero on average currents  to  periodic perturbations \cite{astumian-quantized,westerhoff-86,astumian-91,astumian-90,astumian-jpa05}.
We will call such currents the {\it pump currents}. 
A stochastic pump resembles a {\it quantum pump} \cite{moskalets-review,brower-98,cohen-03,niu-pump,vavilov-review,qpump-exp}. In the latter,
nonzero currents appear under similar circumstances but originate from purely quantum mechanical effects. 
The stochastic pump effect (SPE) was observed in frequency-locked  electronic turnstile devices \cite{turnstile,turnstile-1} and in enzymatic reactions \cite{tsong}. Recently, it was studied experimentally in transport through a conical
nanopore, where  strong pump  current variations  were found as a function of the relative phase of applied voltage signals \cite{pump-exp-07epl}.



\begin{figure}[h]
 \centerline{\includegraphics[width=2.2in]{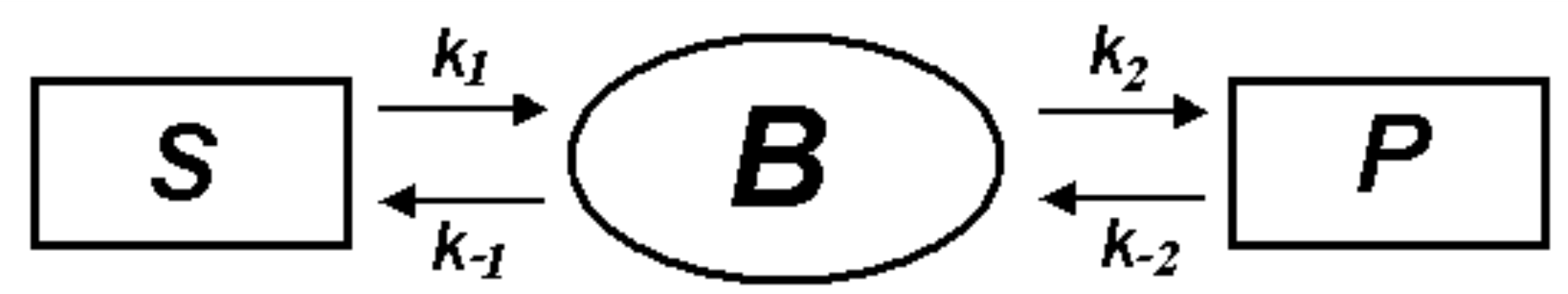}} 
  \caption{\label{system} A simple system demonstrating the stochastic pump effect (SPE).} 
\end{figure}

One of the simplest models
 of the SPE is  illustrated in Fig. \ref{system}.
In this model the central bin  system ${\bf B}$ can have at most one
 particle inside. The bin is connected to the {\it absorbing state} ${\bf S}$ from the left and the absorbing state ${\bf R}$ 
from the right. The term absorbing state  means here that any number of particles can enter this state or leave it. 
Kinetic rates in the model are shown  in Fig. \ref{system}. If the bin is empty, then, with rates $k_1$ or $k_{-2}$, a particle jumps into the bin either from the left or from the right respectively. 
If the bin contains a particle, transitions into the bin are forbidden until this particle escapes to the left or to the right, which it does, respectively,
 with rates $k_{-1}$ and $k_{2}$.
We assume that the kinetic rates are time-dependent,
i.e. they are control parameters, and
we are interested in currents of particles from one absorbing state into the other one. The SPE in this model has been studied in great detail \cite{astumian-03prl,ohkubo-08,sinitsyn-07epl,sinitsyn-07prl,ohkubo-08chem}.  
Even so, due to
the simplicity of the kinetic scheme in Fig.~\ref{system}, the average current  can be studied analytically along with its fluctuations   \cite{sinitsyn-07epl}. 

The model in Fig.~\ref{system} can be used to describe  charge transport through a quantum dot in the Coulomb blockade regime \cite{nazarov-03}, where absorbing states represent conducting leads, and the bin 
represents the quantum dot with at most one electron
 inside. It also serves as a model of
 molecular fluxes through an ion channel connecting two compartments in a living cell \cite{tsong,bezrukov}.

In what follows, we will consider the realization of the model in Fig.~\ref{system} in the  
 enzymatic mechanism of Michaelis-Menten type
 \cite{MM}, which is defined as the following chemical reaction
\begin{equation}
E +S^{\,\,{k_1  \atop \longrightarrow}}_{\,\,  {{\longleftarrow}\atop k_{-1}}}\,\,
 ES ^{\,\,{k_2 \atop \longrightarrow}}_{\,\,{\longleftarrow \atop k_{-2}}}\,E+P,
\label{MM}
\end{equation}
where $S$ and $P$ are called {\it substrate} and {\it product}, and $E$ is an {\it enzyme}
molecule. Here, we assume a situation with a formally infinite number (a {\it sea}) of substrate and product molecules but with only a single enzyme molecule.
Either $S$ or $P$ can combine with the enzyme $E$ to produce an unstable
complex which we will refer to as the {\it bound} state of enzyme.
This complex can then dissociate into either $E$ and $S$ or into $E$ and $P$.
These dissociated states will be referred to as
the {\it free} states. Here it is important to note that, even if $ES$
if formed from $E$ and $P$ (resp. $E$ and $S$), it can freely dissociate into
either $E$ and $P$ or into $E$ and $S$. In terms of the model of Fig. \ref{system}, the
free states are those in which the bin is empty, while the
bound state is that in which the bin is filled.
The sea of substrate molecules is represented by the ${\bf S}$-state and the sea of product molecules is represented by 
the ${\bf P}$-state in Fig.~\ref{system}.

We define {\it the moment generating function} 
for the number of transitions, $n$, in time $T$ from $ES$ (the bin with a particle) into $E+P$ (the absorbing state {\bf P}) by
\begin{equation}
  Z(\chi,T)\equiv e^{S(\chi,T)}=
  \sum_{n=-\infty}^{+\infty} P_n(T)e^{in\chi},
\label{pgf1}
\end{equation}
where $P_n$ is the probability of the event that, by time $T$, there will be $n$ new product molecules created, counting the opposite process with a minus sign.  $S(\chi,T)$ is the {\em  cumulant generating function}
of the number of transitions, because it determines all cumulants of the
 particle flux. For example, the mean $\langle n(T)\rangle$ and the variance
$var(n(T))$ are given, respectively, by
\begin{equation}
\la n(T) \ra = (-i)\left. \frac{\partial S(\chi,T)}{\partial \chi    }\right|_{\chi=0}, \quad {\rm var} (n(T)) = (-i)^2\left. \frac{\partial^2 S(\chi,T)}{\partial \chi^2    }\right|_{\chi=0}. 
\label{cumulants}
\end{equation}
In order to derive the evolution equation
 for the generating function (\ref{pgf1}),  
it is convenient to introduce the generating functions $U_E=\sum_{n=-\infty}^{\infty} P_{nE}e^{in\chi}$ and
$U_{SE}=\sum_{n=-\infty}^{\infty} P_{nSE}e^{in\chi}$, where $P_{nE}$ and $P_{nSE}$ are the
following probabilities:
 $P_{nE}$ is the
probability that, at a given time, the system is
in a free state and the number of product molecules created is $n$, and $P_{nSE}$ is the probability that enzyme is in a bound state and the number of product molecules created is also $n$
\cite{summer-school-08}.
 The master equations for $ P_{nE}$ and $ P_{nSE}$ are then
\begin{equation}
\begin{array}{l}
\frac{d}{dt} P_{nE} = -(k_1+k_{-2})P_{nE} +k_{-1}P_{nSE}+k_2P_{(n-1)SE},\\
\\
\frac{d}{dt} P_{nSE} = -(k_{-1}+k_{2})P_{nSE} +k_{1}P_{nE}+k_{-2}P_{(n+1)E}.
\end{array}
\label{master-2}
\end{equation}
Multiplying (\ref{master-2}) by $e^{i\chi n}$ and summing over $n$ we find 
\begin{equation}
\frac{d}{dt}\left(\begin{array}{l}
 U_{E}\\ 
 U_{SE} 
\end{array}\right)=\hat{H}(\chi,t) \left(\begin{array}{l}
 U_{E}\\ 
 U_{SE} 
\end{array}\right),
\label{master-3}
\end{equation}
where
\begin{equation}
\hat{H}(\chi,t)=\left(
\begin{array}{cc}
-k_1 - k_{-2}   & k_{-1}+k_2 e^{i \chi} \\
k_1+k_{-2}e^{-i\chi}   & -k_{-1}-k_2
\end{array} \right).
\label{hchi}
\end{equation}
If we set $n=0$ at an initial moment $t=0$, then the initial conditions for (\ref{master-3})
are $U_{E}(t=0)=p_E(0)$, and $U_{SE}(t=0)=p_{SE}(0)$, where $p_E(0)$ and $p_{SE}(0)$
 are probabilities that the enzyme is respectively free or in the substrate-enzyme 
complex. Also, note that $Z(\chi,t)=U_{E}(\chi,t)+U_{SE}(\chi,t)$.   
Thus, formally, the moment generating function in (\ref{pgf1}) can be expressed as the following average of the evolution operator
\begin{equation}
Z(\chi,t)= \langle 1\vert \hat{T}\left(e^{\int_{0}^{t}\hat{H}(\chi,t) dt}\right) \vert  p(0) \rangle,
\label{pdf2}
\end{equation}
where $\langle 1 \vert =(1,1)$, where $\vert p(0)\rangle=(p_{E}(0),p_{SE}(0))$ is the vector of initial probabilities of enzyme states, and where $\hat{T}$ is the
time-ordering operator.

A derivation of the adiabatic approximation for (\ref{pdf2}) can 
be found in \cite{sinitsyn-07epl} but the general discussion of section 2 shows that the generating function for a slow cyclic evolution of parameters
is an exponential of the sum of two terms: one geometric and one dynamic,
\begin{equation} 
Z(\chi)= e^{S_{\rm geom}(\chi)+S_{\rm dyn}(\chi)}.
\label{geomph}
\end{equation}
Let $\varepsilon_0(\chi)$ be the
instantaneous eigenvalue of $\hat{H}(\chi,t)$ with the larger real
part,
let $\kappa_{\pm}=k_{\pm1}k_{\pm2}$, let $e_{\pm\chi}=e^{\pm i\chi}-1$  and let $K=\sum_mk_m$ where  $m=-2,-1,1,2$. Then  the {\it dynamic part}
is given by 
\begin{equation}
  S_{\rm dyn}(\chi)=\int_0^{T} dt \varepsilon_0(\chi,t)=-\frac{1}{2} \int_0^{T} dt \left[ K-
  \sqrt{K^2+4(\kappa_+e_\chi+\kappa_-e_{-\chi})}  \right],
\label{Scl}
\end{equation}
The problem of degeneracy of eigenvalues cannot appear here because
the eigenvalue $\varepsilon_0(\chi,t)$ corresponds to the unique steady state of the system.
Here, we note that the vector of kinetic rates ${\bf k}$ depends on $t$ and
describes a contour ${\bf c}$ in the parameter space.
Next, denoting by $\mid u_0(\chi,{\bf k})\rangle$ the eigenvector corresponding
to $\varepsilon_0(\chi,{\bf k})$,   the {\it geometric part} $S_{\rm geom}(\chi)$ is given
by equation
\begin{equation}
  S_{\rm geom}(\chi)=-\oint_{{\bf c}} {\bf A} \cdot d{\bf
    k},\;\,
  A_m = \langle u_0(\chi,{\bf k})|\partial_{k_m}|u_0(\chi,{\bf k})\rangle.
\label{Sgeom}
\end{equation}
There is an obvious analogy between the geometric part of the cumulant generating function (\ref{Sgeom}) and the Berry phase in quantum mechanics (\ref{berry-phase}). 
The geometric phase $S_{\rm geom} (\chi)$ in (\ref{geomph}) was discovered by Sinitsyn and Nemenman in \cite{sinitsyn-07epl}.

If parameters
 $k_1$ and $k_{-2}$ are time dependent, while $k_2$ and
$k_{-1}$ are constants,  then
\begin{equation}
\oint_{{\bf c}} {\bf A} \cdot d{\bf k} = \int\int_{{\bf s_c}} dk_{1}dk_{-2} F_{k_1,k_{-2}},
\label{stokes}
\end{equation}
where ${\bf s_c}$ is a surface whose boundary is
 the contour ${\bf c}$, and
\begin{equation}
F_{k_1,k_{-2}}=\la \frac{\partial u_0 }{\partial k_{1}} \vert \frac{\partial u_0}{\partial k_{-2}} \rangle - 
\la \frac{\partial u_0 }{\partial k_{-2}} \vert \frac{\partial u_0}{\partial k_{1}} \rangle.
\label{berry-curv}
\end{equation}
 $F_{k_1,k_{-2}}$ is the analog of the quantum mechanical
{\it Berry curvature}.  

The following explicit expression for
$F_{k_1,k_{-2}}$ is then obtained \cite{sinitsyn-07epl}
by computing the eigenvectors of the Hamiltonian  
 (\ref{hchi}):
\begin{equation}
F_{k_1,k_{-2}}=\frac{e_{-\chi}(e^{i\chi}k_2+k_{-1})}{[4\kappa_+e_\chi+
4\kappa_-e_{-\chi}+K^2]^{3/2}}.
\label{berry2}
\end{equation}

The generating function (\ref{geomph}) contains information both about average flux and about flux fluctuations. 
To extract the average current from expressions (\ref{Sgeom}) and (\ref{Scl}) one should write the cumulant generating function as a power series in the small parameter $\chi$ and retain the linear part, namely  
\begin{equation}
 S_{\rm dyn} \approx i S^{(1)}_{\rm dyn} \chi +O(\chi^2),\quad S_{\rm geom} =  i \chi \int\int_{{\bf s_c}} dk_{1}dk_{-2} F_{k_1,k_{-2}}^{(1)}+O(\chi^2).
\label{series1}
\end{equation}
Higher order terms in $\chi$  can reveal information about the higher cumulants of stochastic fluxes, while the first order terms coincide, up to the $i\chi$, with the average number of transferred particles.
For the process in Fig.~\ref{system}, such calculations lead to 
the following expression \cite{sinitsyn-07epl} for the mean
 $S \to P$ flux per cycle 
\begin{equation}
  J=J_{\rm geom}+J_{\rm dyn},
\label{JJ}
\end{equation}
\begin{equation}
 J_{\rm geom} =\int \int_{{\bf  s_c}} d^2k 
  \frac{k_2+k_{-1}}{K^3}, 
\label{JJ-10}
\end{equation}
\begin{equation}
 J_{\rm dyn}=\int_0^{T} dt \frac{\kappa_+(t)-\kappa_-(t)}{K(t)}.
\label{JJ-11}
\end{equation}
It turns out that the dynamic contribution $J_{\rm dyn}$ to the current is just the steady state current averaged over time.
Eqs. (\ref{JJ-10}) and (\ref{JJ-11}) show that the geometric contribution $J_{\rm geom}$ to the current has strikingly different properties from the dynamic one. In fact, it does not have an analog in a strict steady state
situation because it is nonzero  only if the contour encloses a finite area in the parameter space, i.e. in order to make it nonzero, at least two parameters should be time-dependent with a phase shift different from $0$ or $\pi$.
 Another interesting property is that this contribution changes sign when  the direction of motion along the contour is reversed.
\begin{figure}[h]
 \centerline{\includegraphics[width=2.2in]{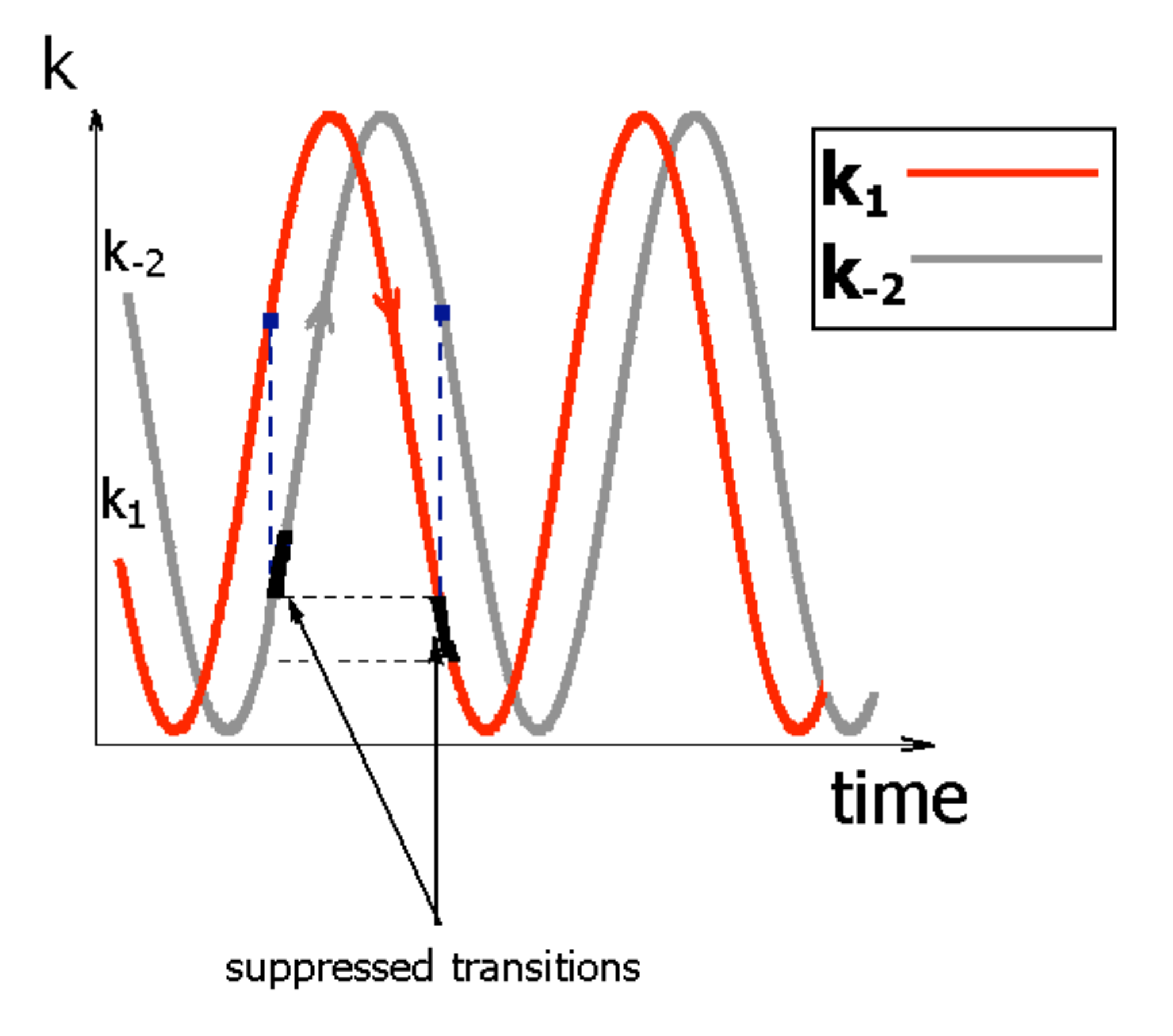}} 
  \caption{\label{PumpShed} Illustration of the shielding mechanism of the SPE.} 
\end{figure}

An intuitive explanation of the phenomenon of $J_{\rm geom}\neq 0$ is illustrated in Fig.~\ref{PumpShed}. 
During an interval of time in which a molecule is bound to the enzyme,
the bin of Fig.~\ref{system} is occupied and the values of $k_1$ and $k_{-2}$
have no effect on the system.
If the left binding rate $k_1$ is higher than the right one $k_{-2}$ during
the upswing of the cycle, then $k_1$ ``shields'' growing values of
$k_{-2}$ from having an effect, while $k_{-2}$ shields decreasing
values of $k_1$ during the downswing. This leads to a phase-dependent
asymmetry which is the source of the geometric pump flux.

It is instructive to compare
 the relation of the geometric phase  (\ref{Sgeom}) to the value given by equation (\ref{eq_lan1}) in the formalism of Landsberg-Ning-Haken, which was discussed in section 3. 
The pump
current (\ref{JJ}) itself can alternatively be derived directly from the  Master Equation for the vector ${\bf p} = (p_{ES},p_E)$ of probabilities of  the enzyme (bin) states. These equations have the form
\begin{equation}
\dot{\bf p} = \hat{H}({\chi=0},{\bf k}) {\bf p}, \quad \dot{N}_{\rm P} =J({\bf p},{\bf k}),
\label{probeq}
\end{equation}
where $N_{\rm P}$ is the average number of product molecules created, and where the current  $J({\bf p},{\bf k})$ is defined by
\begin{equation}
J({\bf p},{\bf k}) = p_{SE}k_2-p_{E}k_{-2}.
\end{equation}
This set of equations has the form (\ref{eq_lan1}), namely, probabilities
  relax to  unique steady state values at given rate constants
${\bf k}$, independently of the additional equation for $N_{\rm P}$. The geometric phase (\ref{Sgeom}), however,
 contains complete information about the stochastic evolution, including geometric contributions to higher cumulants, which cannot be derived from a set of equations in (\ref{probeq}). 
The additional information provided by (\ref{Scl}) and by (\ref{Sgeom}), but not by (\ref{probeq}), is sometimes necessary. For example, it is needed for a complete description of  
 counting statistics of currents  in nanoscale electronic circuits, which were measured experimentally in \cite{sukhorukov-07Nat}.
The higher cumulants are also needed to account for contributions to particle fluxes due to stochastic transitions over potential barriers, as studied in
  \cite{sukhorukov-04prl,sukhorukov-07prl}. Thus the full stochastic treatment of such processes, beyond the formalism of Eqs. (\ref{probeq}), is inevitable,
and the geometric contribution to higher cumulants has important consequences in the theory of such effects.

\section{Non-cyclic geometric phase}
The quantum mechanical Berry phase can be extended to noncyclic evolution \cite{AA,pati-98}. Sinitsyn and Nemenman \cite{sinitsyn-noncyclic} showed that a moment generating function of the form
 (\ref{pdf2}) can 
be partitioned into geometric and dynamic parts even if the parameters change along an open circuit. For evolution during time $\delta t$ we have
\begin{equation}
Z(\chi,\delta t)=e^{S_{\rm geom}(\chi,\delta t)+S_{\rm dyn}(\chi,\delta t)},
\label{mgf3}
\end{equation}
where $S_{\rm dyn}=\int_0^{\delta t} dt \varepsilon_0(\chi,t)$ 
is the quasi-stationary part of the generating function and where
\begin{equation}
S_{\rm geom}=\int_{\bf c} [{\bf P}({\bf k})
-{\bf A}({\bf k})] \cdot d{\bf k},
\label{ssgeom}
\end{equation}
with
\begin{equation} 
{\bf P} =\partial_{{\bf k}} \ln \langle 1 \vert u_{0} \rangle,\quad
{\bf A}({\bf k})=\langle u_0  \vert \partial_{{\bf k}} u_0 \rangle. 
\label{sgeom}
\end{equation}
where $\la 1 \vert =(1,1)$ is the vector with all unit entries.
The noncyclic geometric phase contribution has no analog in a strict steady state regime.
In
general, for a non-cyclic evolution, the
term $-\int_{\bf c}{\bf A}({\bf k})\cdot d{\bf k}$ is not gauge invariant, and the term
$\int_{\bf c}{\bf P}({\bf k})\cdot d{\bf k}$ is a necessary correction to
make it so.
The integral over the additional vector ${\bf P}$  exactly cancels the non-gauge-invariant part of the contour integral of
 ${\bf A}$. The vector ${\bf P}$ introduces the gauge, which  can be derived from proper accounting for the effect of the averaging over the final states of the bin-system at the end of the evolution. 

Since ${\bf P}$ is an exact
differential, it is important only when looking at an evolution along an open path in the parameter space. If the parameter
vector ${\bf k}$
returns to its initial value at the end of the evolution, the expression (\ref{sgeom}) becomes equivalent to the cyclic geometric phase defined in
~\cite{sinitsyn-07epl}. The origin of the gauge invariance of the expression (\ref{ssgeom}) can be traced from the Markovian
 property of the process. The gauge transformation
\begin{equation}
Z(\chi)\rightarrow Z(\chi)e^{S_{\rm prior}(\chi)}.
\label{zzz}
\end{equation}
has the physical meaning of the assumption that additional currents, described by the cumulant generating function $S_{\rm prior}(\chi)$, have passed through the system before the time moment $t=0$.
In Markovian evolution, 
the currents counted after $t=0$ should be independent of the currents counted prior to this moment, which means that the expression (\ref{ssgeom}), describing the currents after $t=0$, is invariant of the gauge transformation 
(\ref{zzz}).

\section{Elimination of fast variables in stochastic processes}
This section is technically more involved than the rest of the review, and can be safely omitted on a first reading. It contains a discussion of the method of stochastic path integral. This is
 a powerful technique for investigating stochastic fluxes in mesoscopic interacting systems. The
discussion in section 5 reduces the problem of computing
 the counting statistics to that of finding eigenvalues and eigenvectors
of a non-Hermitian Hamiltonian. This straightforward approach becomes extremely complicated when applied to
 complex systems with a mesoscopically large phase space. In contrast, the stochastic path integral technique
allows us  to derive moment generating functions of fluxes even in many-body interacting mesoscopic stochastic systems. We advise the reader, who is not familiar with stochastic path integrals,
to study look at the articles
 \cite{kamenev-04,pilgram-03,pilgram-04} before reading this section.
 Sinitsyn and Nemenman \cite{sinitsyn-07prl} demonstrated that this technique can be applied to study 
geometric phases in the evolution of driven mesoscopic stochastic systems. In this section we discuss  another application of geometric phases and the stochastic path integral, namely,
 to the problem of coarse graining stochastic kinetics.

Berry phases often appear in quantum mechanical applications when one attempts to eliminate fast degrees of freedom and reduce a problem to an effective one which includes only slow variables.
Such an approach is known as the  
{\it Born-Oppenheimer approximation} and has been 
 very successful for
describing near-equilibrium properties of most molecules. According to it, initially one solves a much simpler Schr\"odinger equation for electrons,  treating nuclear degrees of freedom as adiabatically slowly changing parameters.
After this, one assumes that the nuclei move on a single
potential-energy surface created by the faster moving
electrons. 
An interesting development
involving this approach was the observation
  that the Berry phase, acquired by electronic wave function in a potential of slowly moving nuclei, influences the
dynamics of slow degrees of freedom  \cite{berry-book,niu-book,BornOp}.

The evolution equations for moment generating functions in stochastic processes are similar to the quantum mechanical Schr\"odinger equation \cite{kamenev-04}.
We already encountered this analogy in section 5.
This mathematical similarity was also used,
 for example, to study counting statistics in electronic transport \cite{pilgram-03,pilgram-04,sinitsyn-07prb}, to estimate over-barrier escape probabilities \cite{sukhorukov-04prl,sukhorukov-07prl}
and to classify stochastic phase
transitions \cite{kamenev-06}. Many of the applications of this approach were restricted to relatively simple systems with only a few interacting species because quantum mechanical equations are usually no simpler to 
investigate than the stochastic ones. 

Recently, a stochastic-quantum analogy was proposed to simulate the behavior of large stochastic networks of biochemical reactions. In general, these networks involve many
different chemical species and reaction types.
 \cite{sinitsyn-08pnas}. One of the difficulties in studying such networks is their {\it stiffness}, i.e.  strong time-scale separation of various processes in a network. 
In its application to stochastic processes, the Born-Oppenheimer approximation 
rigorously captures statistical characteristics of chemical processes at  coarse-grained scales \cite{sinitsyn-08pnas}.  


In this section, we present a simple example of how a reduction of a model can be achieved and how geometric phases influence the evolution of slow variables.
For this purpose we again consider the
Michaelis-Menten type of conversion of $S$ into $P$ via creation of a substrate-enzyme complex, which we studied in section 5. Now, however,  we  assume 
that numbers of substrate and product molecules ($N_{\rm S}$ and $N_{\rm P}$ respectively) are independent dynamic variables,  subject to all the conservation laws imposed
 by the given kinetic scheme.
We also assume that the enzyme-substrate complex is created from substrate and product molecules at kinetic rates which are respectively  proportional to the absolute numbers 
 $N_{\rm S}$ and $N_{\rm P}$ of these molecules. We then have to consider the following four reactions 
\begin{enumerate}
\item forward substrate-enzyme complex formation, $S+E \rightarrow SE$,
  with rate $k_1N_{\rm S}$;
\item backward substrate-enzyme complex formation, $P+E \rightarrow SE$,
  with rate $k_{-2}N_{\rm P}$;
\item complex backward decay, $SE \rightarrow S+E$, with rate
  $k_{-1}$;
\item product emission $SE \rightarrow E+P$, with rate $k_{2}$.
\end{enumerate}
Since we have only
  one enzyme molecule but $N_{\rm S},N_{\rm P} \gg 1$,  it takes many identical steps for the enzyme molecule  to convert a substantial number of substrates into products. 
This creates a time-scale separation, which can be used to reduce the model to an effective process 
\begin{equation}
S\rightarrow P.
\label{effproc}
\end{equation}
Our goal is to find the statistical characteristics of the coarse-grained reaction (\ref{effproc}).
Let $T_{S\rightarrow P}$ be the time it takes an enzyme molecule to convert a substrate molecule to a product molecule. We choose a time scale $\delta t$ which is much larger than $T_{S\rightarrow P}$,
but much smaller than $N_ST_{S\rightarrow P}$.
 Suppose that we are looking for the moment generating function of the number  $n_{\rm P}$ of product molecules created during a relatively long time $T\gg \delta t$:
\begin{equation}
 {\mathcal Z}(\chi_{\rm C})=e^{{\mathcal S}(\chi_{\rm C})}=
 \sum_{n_{\rm P}=-\infty}^{\infty} P(n_{\rm P}|T)e^{in_{\rm P}\chi_{\rm C}}.
\label{MGF-1}
\end{equation}
Here we introduce an additional index $C$ to mark a counting parameter in (\ref{MGF-1}) in order to distinguish it from other variables which will appear in the following calculations.
As  noted in \cite{pilgram-03,kamenev-04}, if one knows the statistical properties of fluxes at time scales $\delta t$ then the moment generating function at larger time scales can be written in the form of a stochastic
 path integral.
We will derive the stochastic path integral representation of the generating function in (\ref{MGF-1}).

In Sections 5-6 we already found the full counting statistics of the fluxes in Michaelis-Menten model with slowly time dependent parameters. To apply it to our model we should  redefine kinetic rates as $k_1\rightarrow k_1N_{\rm S}(t)$ and
$k_{-2}\rightarrow k_{-2}N_{\rm P}(t)$. This does not solve our problem completely, because at this stage we do not know the explicit time dependence of $N_{\rm S}(t)$ and $N_{\rm P}(t)$, which we regard as slow but dynamic variables. 
Let us partition the time line  into intervals $(t_m,t_m+\delta t)$ of  durations
$\delta t$, where
$t_m=m\delta t$,
 and
let $  \delta n_{\rm P}(t_m)$ be the relative change in the number of product molecules during these time intervals.
We chose $\delta t$ sufficiently small, so that $ \la \delta n_{\rm P}(t_m) \ra /N_{\rm P/S}(t_m) \ll 1$ while the absolute change is large $\la \delta n_{\rm P}(t_m) \ra \gg 1$.
The probability
distributions of the $ \delta n_{\rm P}(t_m) $ are then given by the inverse Fourier
transforms of the corresponding moment generating functions in Section 6:
 \begin{equation}
 P(\delta n_{\rm P}(t_m)) =  \int_{-\pi}^{\pi} \frac{d \chi (t_m)}{2\pi}
 \exp \left({-i\chi(t_m) \delta n_{\rm P}(t_m) + S_{\rm MM} (\chi(t_m),\delta t) } \right). 
\label{Pij}
\end{equation}
Here, $S_{\rm MM}=S_{\rm geom}+S_{\rm dyn}$ is the cumulant generating function of the number of new product molecules which were generated during time $\delta t$. Treating $N_{{\rm S/P}}(t)$ as slow time-dependent parameters,
one finds that $S_{\rm MM}$ is given by Eq. (\ref{mgf3}).

The moment generating function of the number of product molecules created during a 
large time interval $(0,T)$ 
is given by the sum over all possible paths in the space of dynamic variables $N_{\rm S}(t_m)$, $N_{\rm P}(t_m)$ and $\delta n_{\rm P}(t_m)$,
weighted by the probabilities (\ref{Pij}) and by delta-functions, which are responsible
for conservation of the number of molecules,
\begin{equation}
 \delta_{N_{\rm S}}^m=\delta (N_{\rm S}(t_{m+1})-N_{\rm S}(t_m) + \delta n_{\rm P} (t_m) ),
\end{equation}
\begin{equation}
 \delta_{N_{\rm P}}^m=\delta (N_{\rm P}(t_{m+1})-N_{\rm P}(t_m) - \delta n_{\rm P} (t_m) ).
\end{equation}
We use delta-functions  instead of Kronecker symbols because the assumption $N_{\rm S/P},\delta n_{\rm P} (t_m)  \gg 1$ allows to treat  $N_{\rm S/P}$ and $\delta n_{\rm P}$ as continuous variables.
Using that $n_{\rm P}=\sum \limits_{m=1}^{T/\delta t} \delta n_{\rm P}(t_m)$, the moment generating function of this
number is a discrete path integral
\begin{eqnarray}
\noindent Z(\chi_{\rm C}) \equiv \langle e^{i\chi_{\rm C}  n_{\rm P}}\rangle =\nonumber \\
\prod_m \int dN_{\rm S}(t_m) dN_{\rm P}(t_m)  d(\delta n_{\rm P}(t_m)) 
P[\delta n_{\rm P}(t_m)] 
e^{i\chi_C  \delta n_{\rm P}(t_m)}\delta_{N_{\rm S}}^m \delta_{N_{\rm P}}^m.
\label{path1}
\end{eqnarray}
We  rewrite the delta-functions in  (\ref{Pij}) as integrals over oscillating exponents
$$
\delta_{N_{\rm S/P}}^m =
\frac{1}{2\pi} \int_{-\pi}^{+\pi} d\chi_{\rm S/P}(t_m)\exp \left(i\chi_{\rm S/P}(t_m)[N_{\rm S/P}(t_{m+1})-
N_{\rm S/P}(t_m) \pm \delta n_{\rm P} (t_m)]\right),
$$
and substitute the result into (\ref{path1}). After this, 
integrals over $\delta n_{\rm P}(t_m)$ produce new delta-functions, which 
are easily removed by integration over $\chi(t_m)$, leaving us only
with a path integral over the slow variables $N_{\rm S/P}(t_m)$ and $\chi_{\rm S/P}(t_m)$.  Taking a continuous limit, we have $N_{\rm S/P}(t_{m+1})-N_{\rm S/P}(t_{m}) \rightarrow \dot{N}_{\rm S/P}(t)dt$,
and we can rewrite the moment generating function as a path integral involving only slow variables:
\begin{equation}
\langle e^{i\chi_C n_{\rm P}}\rangle = 
\int DN_{\rm S}(t) \int DN_{\rm P}(t) \int D \chi_{\rm S} (t) \int D \chi_{\rm P} (t) e^{S(\chi_{\rm C},T)} ,
\label{path33}
\end{equation}
where $S(\chi_{\rm C},T)$ can be partitioned into  dynamic and
geometric parts 
\begin{equation}
  S(\chi_{\rm C},T) =  S^{\rm dyn}(\chi_{\rm C},T)+ S^{\rm geom}(\chi_{\rm C},T),
\label{fcs3}
\end{equation}
such that
\begin{equation}
  S^{\rm geom}=-\int_{{\bf c}} [ A_{N_{\rm S}} dN_{\rm S} + A_{N_{\rm P}}dN_{\rm P}+
A_{\chi_{\rm S}} d\chi_{\rm S} +A_{\chi_{\rm P}}d\chi_{\rm P}],
\label{sgeom55}
\end{equation}
\begin{equation}
  A_x= \langle u_0(\chi_{\rm S}-\chi_{\rm P}+\chi_C)|\partial_{x}|u_0(\chi_{\rm S}-\chi_{\rm P}+\chi_{\rm C})\rangle,
\label{sgeom6}
\end{equation} 
\begin{equation}
  S^{\rm dyn}=\int_{0}^{T}dt\, [ i\chi_{\rm S} \dot{N}_{\rm S}+
i\chi_{\rm P} \dot{N}_{\rm P}+H_{\rm MM}(N_{\rm S},N_p,\chi_{\rm S}-\chi_{\rm P}+\chi_{\rm C}) ],
\label{scl5}
\end{equation}
where ${\bf c}$ is the contour of the trajectory in 
the slow parameter space,
 $x$ belongs to the set $\{N_{\rm S},N_{\rm P},\chi_{\rm S},\chi_{\rm P} \}$
and $H_{\rm MM}$ plays the role of the effective Hamiltonian
$$
H_{\rm MM}=  \frac{ K- \sqrt{K^2+4[N_{\rm S}k_1k_2(e^{i(\chi_{\rm S}-\chi_{\rm P}+\chi_{\rm C})}-1)+
N_{\rm P}k_{-1}k_{-2}(e^{-i(\chi_{\rm S}-\chi_{\rm P}+\chi_{\rm C})}-1)] }  }{2},
$$
where $K\equiv k_1N_{\rm S}+k_{-2}N_{\rm P}+k_{-1}+k_2$.
This Hamiltonian
is different from what one would expect if the conversion of $S$ into $P$ were
a Poisson process, which reflects the non-Poisson nature of enzyme mediated
fluxes. 
The geometric part of the action (\ref{sgeom55}) is the result of the
pump fluxes.

The path integral  (\ref{path33}) is a formal solution of 
the problem of removing of fast degrees of freedom: it expresses the moment generating function in terms of only
slow variables $N_{\rm S},N_{\rm P},\chi_{\rm S},\chi_{\rm P}$ and does not
depend on the degrees of freedom of the enzyme. 

Since the averages
 $\la N_{\rm S} \ra$ and $\la N_{\rm P} \ra$ are
 assumed large, one can use the saddle point 
solution of the path integral 
to derive semiclassical equations of motion for slow variables.
Varying the action results in four coupled 
differential equations
\begin{eqnarray}
i\dot{ N}_{\rm S} &= -\frac{\partial H_{\rm MM}}{\partial \chi_{\rm S}}-
iF_{\chi_{\rm S},N_{\rm S}} \dot{N}_{\rm S} - i F_{\chi_{\rm S} N_{\rm P}}\dot{N}_{\rm P}, \\
i\dot{ N}_{\rm P} &= -\frac{\partial H_{\rm MM}}{\partial \chi_{\rm P}}+
iF_{\chi_{\rm S},N_{\rm S}} \dot{N}_{\rm S} + i F_{\chi_{\rm S} N_{\rm P}}\dot{N}_{\rm P}, \\
i\dot{ \chi}_{\rm S} &= \frac{\partial H_{\rm MM}}{\partial N_{\rm S}}-
iF_{\chi_{\rm S},N_{\rm S}} \dot{\chi}_{\rm S} - i F_{\chi_{\rm S} N_{\rm P}}\dot{\chi}_{\rm P} +iF_{N_{\rm S},N_{\rm P}}\dot{N}_{\rm P},\\
i\dot{ \chi}_{\rm P} &= \frac{\partial H_{\rm MM}}{\partial N_{\rm P}}+
iF_{\chi_{\rm S},N_{\rm S}} \dot{\chi}_{\rm S} + i F_{\chi_{\rm S} N_{\rm P}}\dot{\chi}_{\rm P} -iF_{N_{\rm S},N_{\rm P}}\dot{N}_{\rm S},
\label{eqmotion1}
\end{eqnarray}
where
$F_{x_1,x_2}=i(\partial A_{x_2}/\partial x_1-\partial A_{x_1}/\partial x_2)$ and
we used  that $A_{x_m}$  depends on $\chi_{\rm S}, \chi_{\rm P}$ 
via the combination $(\chi_{\rm S}-\chi_{\rm P})$ which leads to the relations
$F_{\chi_n,\chi_m}=0$, $F_{\chi_{\rm S},N_{\rm S}}=-F_{\chi_{\rm P},N_{\rm S}}$ and $F_{\chi_{\rm S},N_{\rm P}}=-F_{\chi_{\rm P},N_{\rm P}}$.  
The boundary conditions can be also influenced by geometric phases and should be derived by the method used in \cite{kamenev-04,pilgram-03}.  

For $\chi_{\rm C}=0$ there is a solution of equations (\ref{eqmotion1})
such that $\chi_{\rm S}=\chi_{\rm P}=0$. $N_{\rm S}$, $N_{\rm P}$ then satisfy coupled equations, which are known to
coincide with the mean field equations \cite{pilgram-03}
\begin{eqnarray}
i\dot{ N}_{\rm S} &=&
-i\Omega_{\chi_{\rm S},N_{\rm S}} \dot{N}_{\rm S} - i \Omega_{\chi_{\rm S} N_{\rm P}}\dot{N}_{\rm P}
 -\frac{\partial H_{\rm MM}}{\partial \chi_{\rm S}}|_{\chi_{\rm S}=\chi_{\rm P}=\chi_{\rm C}=0}, \\
i\dot{ N}_{\rm P} &=&
i\Omega_{\chi_{\rm S},N_{\rm S}} \dot{N}_{\rm S} + i \Omega_{\chi_{\rm S} N_{\rm P}}\dot{N}_{\rm P} 
 -\frac{\partial H_{\rm MM}}{\partial \chi_{\rm P}}|_{\chi_{\rm S}=\chi_{\rm P}=\chi_{\rm C}=0},
\label{eqmotion2}
\end{eqnarray}
where $\Omega_{x_1,x_2}=F_{x_1,x_2}\vert_{\chi_{\rm S}=\chi_{\rm P}=\chi_C=0}$.
Explicitly, for our model one can find that 
\begin{eqnarray}
\Omega_{\chi_{\rm S},N_{\rm S}}&=-k_1(k_2+k_{-1})\frac{(k_2+k_{-2}N_{\rm P})}{K^3},\\
\Omega_{\chi_{\rm S},N_{\rm P}}&=-k_{-2}(k_2+k_{-1})\frac{(k_2+k_{-2}N_{\rm P})}{K^3}.
\label{om}
\end{eqnarray}
If substrate and product have no other dynamics
than the conversion into each other via the $ES$ complex,
then $\dot{N}_{\rm P}=-\dot{N}_{\rm S}$ and
$$
\frac{\partial H_{\rm MM}}{\partial \chi_{\rm S}}\vert_{\chi_{\rm S}=\chi_{\rm P}=\chi_C=0}
=-\frac{\partial H_{\rm MM}}{\partial \chi_{\rm P}}\vert_{\chi_{\rm S}=\chi_{\rm P}=\chi_C=0}
=i(k_1k_2N_{\rm S}-k_{-1}k_{-2}N_{\rm P})/K,
$$
so, the evolution equation (\ref{eqmotion2}) for $N_{\rm P}$ becomes
\begin{eqnarray}
\dot{N}_{\rm P} &=&\frac{(k_1k_2N_{\rm S}-k_{-1}k_{-2}N_{\rm P})}{K} - \nonumber \\
&-&\frac{(k_2+k_{-1})(k_1\dot{N}_{\rm S}+k_{-2}\dot{N}_{\rm P})(k_2+k_{-2}N_{\rm P})}{K^3}.
\label{eqmotion3}
\end{eqnarray}
The first term in (\ref{eqmotion3}) is  the usual quasi-steady-state prediction for the substrate-product conversion rate, and the second
term is a correction due to the geometric phase contribution to the effective action.
We note again that Eqs. (\ref{eqmotion1}) contain information both about average fluxes and their fluctuations, while the result (\ref{eqmotion3}) 
is equivalent to the mean-field prediction for the average number of $N_{\rm P}$. Substituting the solution of (\ref{eqmotion1}) into (\ref{fcs3}), one obtains the 
full counting statistics of the number $n_{\rm P}$ of product molecules created.

Terms resembling the geometric phase corrections in (\ref{eqmotion1}) also appear naturally in  variational approaches to chemical kinetics \cite{ohkubo-wp}.
Equations of motion, such as
(\ref{eqmotion1}) are known in 
 condensed matter physics, where similar Berry phase terms
 lead to distinct effects, such as the anomalous and the spin Hall effects 
\cite{sinitsyn-rev,murakami-she,sinova-she}. Generally, the geometric phase correction in 
(\ref{eqmotion3}) is  smaller than the quasi-steady state part.
However, there are situations when it can be 
important due to its specific symmetries \cite{sinitsyn-noncyclic}.

\section{Driven limit cycle}
An interesting geometric phase was found by Kagan {\it et al} \cite{kagan-91} in dissipative systems evolving to a limit cycle.
In addition to a geometric phase that appears after the elimination of quickly decaying modes, they found also a geometric phase which arises from the nontrivial
topology of the limit cycle itself. Consider the following evolution equation
\begin{equation}
\frac{d\phi}{dt} = \Omega(\phi,{\bf \mu}),
\label{ev1}
\end{equation}
where $\Omega(\phi)= \Omega(\phi+2\pi)$ is the instantaneous frequency and ${\bf \mu}$ is a vector of internal parameters, which is slowly and periodically time-dependent.
Since the evolution is periodic, we can regard $\phi$ as a phase.
Introducing the new variable
\begin{equation}
\theta(\phi,{\bf \mu}) = \int_0^{\phi}\frac{\omega({{\bf \mu}})}{\Omega(\phi',{\bf \mu})} d\phi',
\label{theta1}
\end{equation}  
which is a rescaled version of $\phi$, defined so that $\theta$ evolves at the constant rate  
\begin{equation}
\omega({\bf \mu}) = \left( \frac{1}{2\pi} \int_0^{2\pi}\frac{1}{\Omega(\phi',{\bf \mu})} d\phi' \right)^{-1}
\label{omega1}
\end{equation}
when $\mu$ is fixed,
Kagan {\it et al}  showed that for  adiabatic cyclic evolution of ${\bf \mu}$ the phase (\ref{theta1}) becomes the sum of dynamic and geometric parts,
\begin{equation}
\theta (T) = \theta_{\rm dyn}(T)+\theta_{\rm geom}(T), 
\label{theta2}
\end{equation}
where $T$ is the period of the adiabatic evolution of parameters, 
\begin{equation}
\theta_{\rm dyn}(T)=\int_0^Tdt \omega ({\bf \mu}(t)),
\label{theta_dyn}
\end{equation}
and 
\begin{equation}
\theta_{\rm geom}(T)=\oint {\bf A \cdot} d{\bf \mu} ,
\label{theta_geom}
\end{equation}
where
\begin{equation}
{\bf A}=\int_0^{2\pi}\frac{d\phi}{2\pi} \left[ \frac{\omega ({\bf \mu})}{\Omega(\phi,{\bf \mu})} \partial_{{\bf \mu}} \theta(\phi,{\bf \mu})) \right].
\label{conn}
\end{equation}

One can look at the geometric phase (\ref{theta_geom}) from the point of view of the stochastic path integral representation, which was discussed in section 7.
As in the derivation of the stochastic path integral, one can promote
the evolution (\ref{ev1}) to a Hamiltonian flow by introducing a variable $\Lambda$, which is canonically conjugate to $\phi$, with the Hamiltonian
\begin{equation}
H(\Lambda,\phi)=\Lambda \Omega(\phi,{\bf \mu}).
\label{ham1}
\end{equation}
The phase evolution (\ref{ev1}) then follows from the canonical equation
\begin{equation}
\frac{d\phi}{dt}=\frac{\partial H}{\partial \Lambda}.
\label{ev3}
\end{equation} 
 Sinitsyn and Ohkubo \cite{sinitsyn-08jpa} showed that, in this Hamiltonian evolution,  $\theta_{\rm geom}$ becomes a Hannay angle \cite{hannay-85}, 
which is a geometric phase in
classical mechanics that is  responsible for the rotation of the Foucault pendulum, and many other subtle effects.

\section{Thermodynamic constraints and geometric phases}
Geometric approach to classical thermodynamics began with Josiah Willard Gibbs' pioneering work,
 called ``Graphical Methods in the Thermodynamics of Fluids'' \cite{gibbs}. 
With the development of this point of view,
it has become
 possible to formulate classical equilibrium and near-equilibrium thermodynamics in terms of the theory of metric spaces and the vector geometry  \cite{geometry-td}. 

Many applications of  geometric phases in stochastic kinetics involve perturbation of systems initially in thermodynamic equilibrium. These include the response of molecular motors or mesoscopic electronic 
circuits to periodic evolution of parameters. 
The laws of thermodynamics impose  constraints on the kinetic rates which guarantee that the Boltzmann distribution at a given temperature 
describes the equilibrium state. These constrains also
have important consequences for geometric phases.

Were it not for geometric phases, a thermodynamic system with adiabatically changing parameters would have no average current.
When detailed balance conditions are imposed on kinetic rates at any moment of time, a strict quasi-steady state 
approximation predicts exactly zero fluxes on average in response to adiabatically slow perturbations. Hence, the geometric phase is the only mechanism that can be responsible for  nonzero currents 
in such systems.
In this section, we discuss several examples in support of this conclusion.

\subsection{Reversible ratchet}
\begin{figure}[h]
 \centerline{\includegraphics[width=2.4in]{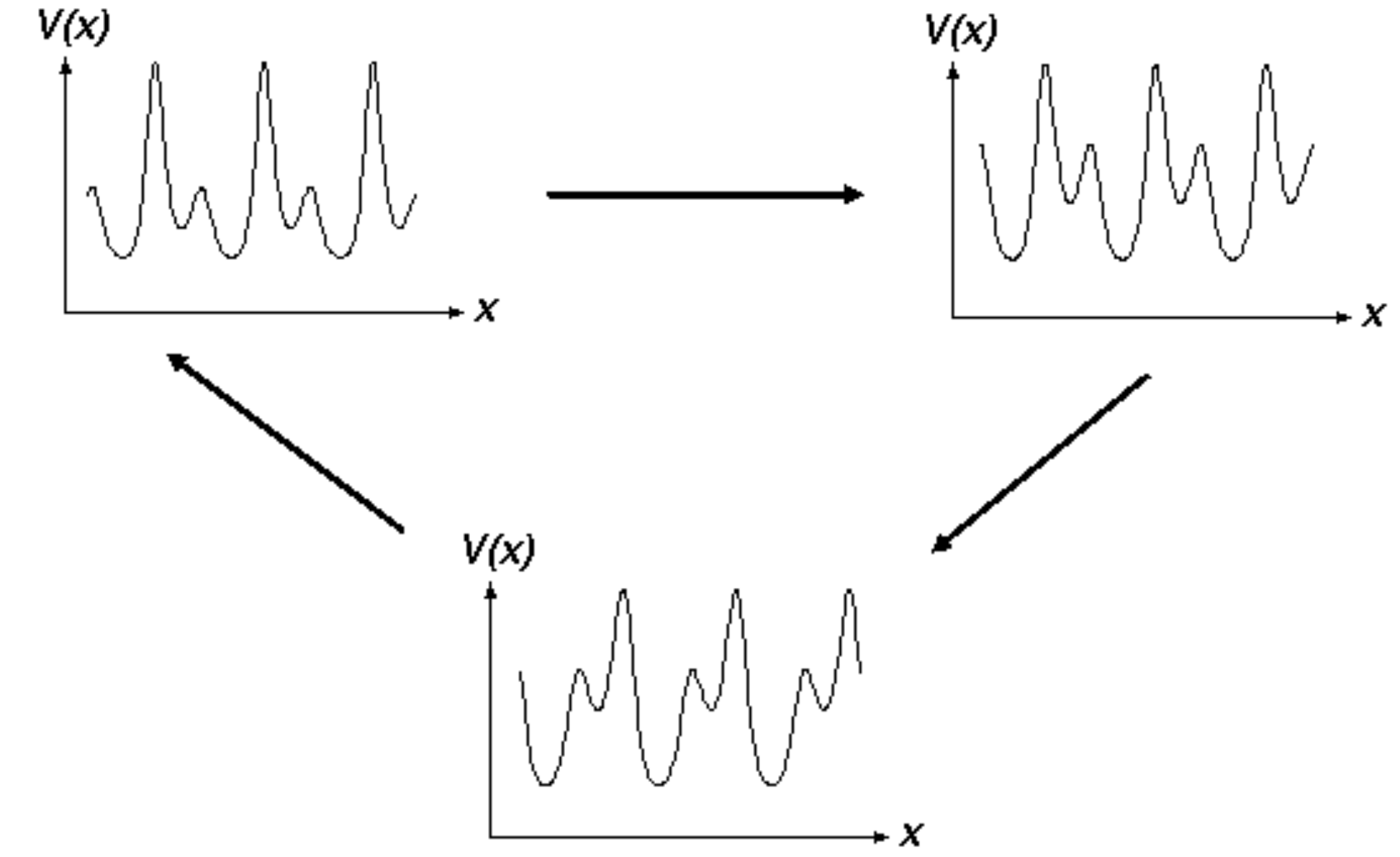}} 
  \caption{\label{ratchet} Snapshots of a ratchet potential at three stages of its evolution.} 
\end{figure}

{\it Ratchets} are systems, where particles diffuse in a potential, which is periodic both in space and time, i.e. $V(x,t)=V(x+L,t)=V(x,t+T)$. 
 The particle distribution  $\rho(x,t)$ satisfies the Fokker-Planck equation, which predicts that for a fixed
potential profile, $\rho(x,t)$ relaxes to the Boltzmann distribution $\rho(x) = C e^{-V(x)/k_BT}$, where $C$ is a normalization constant. 
An example of a ratchet is shown in Fig.~\ref{ratchet}.
The ratchet working in the regime of adiabatically slow evolution of the potential profile is called the
 {\it reversible ratchet} because currents in this limit change sign when the potential profile changes with time in a reversed order.  
Thus, the reversible ratchet is a simple example of a device working near thermodynamic equilibrium.

Currents in a discrete version of a reversible ratchet were studied by Markin and Astumian
\cite{markin-astumian}.
The continuous model \cite{parrondo-98} was studied by Parrondo, who derived the explicit expression for the current of particles in such a system in the limit of adiabatically slow changes of the potential.
The paradox is that, for adiabatically  slow changes, one can expect that the 
particle distribution would have enough time to converge to an instantaneous equilibrium distribution, i.e. it is expected to have a form 
$\rho(x,t) \approx C(t) e^{-V(x,t)/k_BT}$. Such a varying Boltzmann distribution does not predict any current on average in the system at any moment of time, while apparently the solution of the problem predicts a nonzero current.

Sinitsyn and Nemenman \cite{sinitsyn-07prl} explored this model from the point of view of the stochastic path integral representation of the  moment generating function of particle currents $Z(\chi,T)$. They showed that 
\begin{equation}
Z(\chi,T)=e^{i\chi \oint_{{\bf c}} {\bf A}({\bf k}) \cdot d{\bf k} +O(\chi^2)},
\label{ratchet2}
\end{equation}
where ${\bf k}$ is the vector of parameters  controlling the shape of $V(x)$, and ${\bf c}$ is the contour in this parameter space.
 According to (\ref{ratchet2}), the geometric phase is not zero, and contributes to the linear term in $\chi$ of the moment generating function.
This  confirms that the
current in an adiabatic reversible ratchet is nonzero and arises purely geometrically. It can be totally controlled by choosing a proper contour in the space of potential shapes.

An interesting observation about this effect was made by Shi and Niu \cite{shi}, who showed that this current can be quantized and that this quantization can be related to a
 Chern number of a Bloch band related to 
the periodic potential.

\subsection{Geometric phases and fluctuation-dissipation relations.}

 The adiabatic SPE appears
when two time-dependent periodic perturbations are applied. From the discussion in section 5, it follows that the average 
flux of the ``charge'', pumped by an application of infinitesimal cyclic  external fields
$ h_B (t)$ and $h_C(t)$, is proportional to the area inside the contour of parameter evolution. 
We also showed that the pump current reverses its sign when a system is moving along the same
contour but in the opposite direction. This situation can be expressed as the following law,
\begin{equation}
\delta q= F_{BC}  dh_B \wedge dh_C,
\label{nonlin}
\end{equation}
where $\delta q$ is the average of the flux that  passes through the system during one cycle of the adiabatic periodic evolution, $dh_B \wedge dh_C$ is the
 infinitesimal directed area enclosed by the contour
in the  space of control parameters, and $ F_{BC}$ is the proportionality coefficient which, up to an $i\chi$-factor, coincides with the  part of the Berry curvature linear in $\chi$. 
 
According to the  Fluctuation-Dissipation Theorem \cite{fluctuation-book}, some transport coefficients near the point of thermodynamic equilibrium  can be expressed 
in terms of correlation functions at the equilibrium point.
Relations of this kind tell us a lot about the coefficient $F_{BC}$ \cite{ cohen-03,astumian-pre09}. 
The quantum version of the adiabatic pump effect can be quantified using the Kubo formula \cite{cohen-03}.
One can derive the analogous result for a classical stochastic pump, operating near thermodynamic equilibrium, using a classical version of linear response theory \cite{lax-fdt}.

Assume that the variables $B$ and $C$  coupled to the fields $h_B$ and $h_C$ in expression for the thermodynamic potential are invariant under time reversal.
Then at the thermodynamic equilibrium, with fixed $h_B$ and $h_C$, all currents are zero on average, that is
\begin{equation}
\langle  J(t) \rangle_{h_B,h_C}=0,
\label{av1}
\end{equation}
where $\langle \ldots \rangle_{h_Bh_C}$ denotes the average over the equilibrium distribution at given values of $h_B$ and $h_C$.
If we start at equilibrium and increase $h_B$, $h_C$ by small amounts $\delta h_B$ and $\delta h_C$, then $h_{\alpha}(t)=h_{\alpha}(0)+\delta h_{\alpha} \theta (t)$, where ${\alpha}=B,C$.
 According to linear response theory, the current at a moment $t>0$ is given by
\begin{equation}
 J(t)=\delta J_{B}(t)+\delta J_{C}(t),
\label{curr}
\end{equation} 
where
\begin{equation}
\delta J_{{\alpha}}(t)=2i\int_{0}^{t}dt' \chi_{{J {\alpha}}}'' (t-t') \delta h_{\alpha},\label{jab}
\end{equation}
and $\chi_{{J\alpha}}''(\omega)$ is the 
response function \cite{fluctuation-book}. Letting $\chi_{J,\alpha}''(\omega)$ be the Fourier transform of $\chi_{{J , {\alpha}}}'' (t)$, we then have
\begin{equation}
\delta J_{{\alpha}}(t)=
 \int \frac{d\omega}{2\pi} \frac{2 \delta h_{\alpha} \chi_{{J {\alpha}}}'' (\omega)}{\omega } e^{i\omega t}. 
\label{jab1}
\end{equation}
Here,  we have used the fact that $J$ and $B$ (resp. $C$)
 have different time reversal properties so that the static susceptibility \cite{fluctuation-book} is identically zero, i.e.
\begin{equation}
\int \frac{d\omega}{2\pi}  \frac{ \chi_{{J {\alpha}}}'' (\omega)}{\omega } =0.
\label{static}
\end{equation}
The classical {\it Fluctuation-Dissipation Theorem}  \cite{fluctuation-book} states that 
\begin{equation}
\chi_{J\alpha}''(\omega)=\frac{\beta \omega}{2} S_{{J \alpha}}(\omega), \quad \beta =1/k_{\rm B}T,
\label{FDT}
\end{equation}
where $S_{J\alpha}(\omega)$ is the Fourier transform of the correlator 
\begin{equation}
S_{{J \alpha}}(t-t')=\langle J(t) \alpha(t') \rangle,\quad \alpha=B,C
 \label{sss}
\end{equation}
at equilibrium.
Substituting (\ref{FDT}) and (\ref{sss}) into (\ref{jab1}) and integrating over time, the total flux $\delta q$ passed due to a perturbation is
\begin{equation}
\delta q=Q_{JB} \delta h_B+Q_{JC} \delta h_C,
\label{dqab}
\end{equation}
where
\begin{equation}
Q_{JB}=\beta \int_0^{\infty} dt S_{JB}(t), 
\label{qab}
\end{equation}
\begin{equation}
Q_{JC}=\beta \int_0^{\infty} dt S_{JC}(t). 
\label{qac}
\end{equation}
Note that both $Q_{JB}$ and $Q_{JC}$ are completely determined by the equilibrium correlation functions for nonzero $B$ and $C$.
If we imagine that the adiabatic evolution in the  space of control parameters consists of  small perturbations, as above, 
then the total transferred charge after one cycle  is
\begin{equation}
\delta q=\oint_{{\bf c}} \{ Q_{JB}dh_B+Q_{JC}dh_C \} = \int\int_{{\bf s_c}} dh_B \wedge dh_C F_{BC},
\label{dqab}
\end{equation}
where ${\bf c}$ is the contour in the  space of control parameters and $F_{BC}$ is the transport coefficient that we have been looking for. According to Stokes' theorem
\begin{equation}
F_{BC}(h_B,h_C)=\frac{\partial Q_{JC}}{\partial h_B} - \frac{\partial Q_{JB}}{\partial h_C}.
\label{fabc}
\end{equation}

Eq. (\ref{fabc}) shows that the pump transport coefficient can be expressed as the circulation, in the
space of controlled parameters, of a vector ${\bf Q}$ whose
components  are the values of the correlators (\ref{qab},\ref{qac}) at equilibrium.  Such relations indicate that the ability of a system to perform as
 a stochastic pump in response to periodic perturbations, can be inferred,
 in principle, from system properties at thermodynamic equilibrium. 
This can lead to  simplifications during numerical or perturbative analysis of molecular motor operations, because
the equilibrium properties are relatively easy to investigate. For example,
calculations of free energy landscapes are achievable for complex biological molecules such as the kinesin molecular motor \cite{FE-kinesin}.

\subsection{Beyond adiabatic and perturbative limits}

\begin{figure}
\includegraphics[width=9.0 cm]{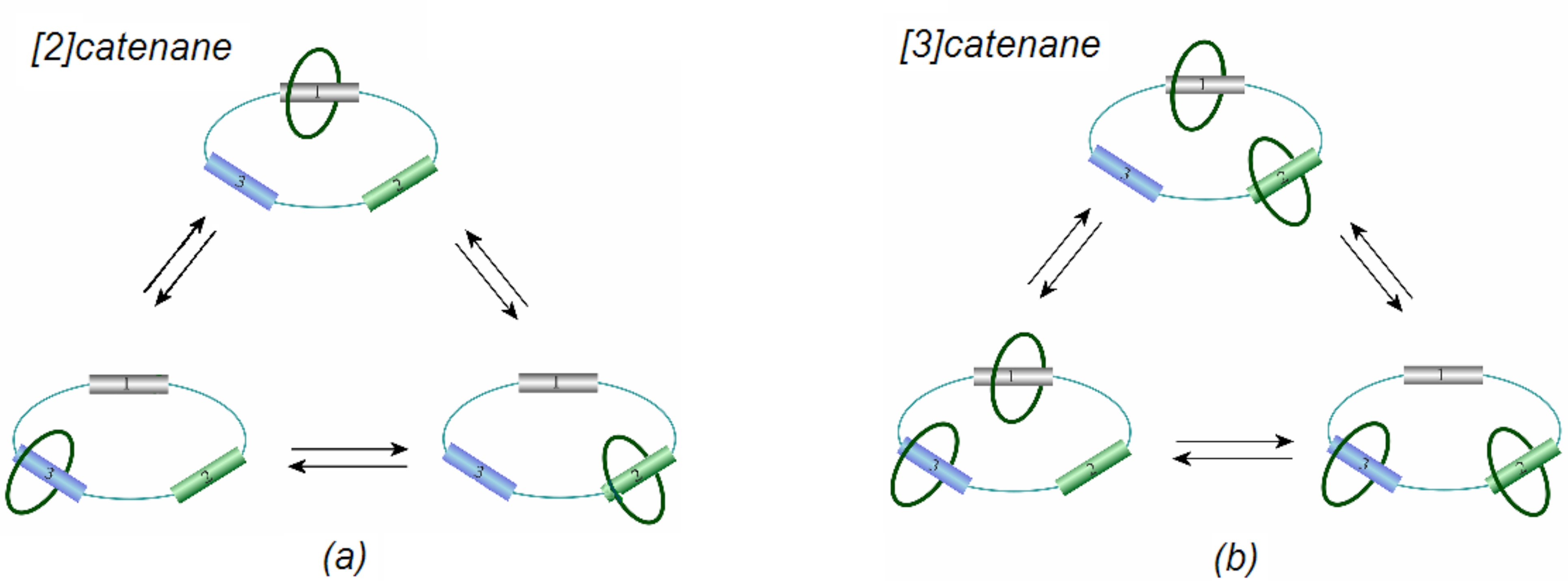}
\centering
\caption{\label{catenane} Topology and metastable states of (a) [2]catenane and (b) [3]catenane molecules. Smaller rings are capable to perform directed rotational motion around the larger ring in response to
external periodic driving. In [3]catenanes, unlike [2]catenanes,  directed motion can be achieved even when only coupling strengths of smaller rings to the three stations on the larger ring are controlled.}
\end{figure}
So far, we have only discussed the systems, which are driven adiabatically slowly. 
Quantum mechanical Berry phase can be generalized to the case of non-adiabatic evolution \cite{AA}. Following this analogy, Ohkubo showed that 
geometric phases in stochastic kinetics also can be considered in the non-adiabatic regime \cite{ohkubo-08}. 

Recently, a number of exact results in the theory of the stochastic pump effect were derived which are valid in 
nonperturbative and nonadiabatic regimes \cite{jarzynski-nopump, sinitsyn-08prl}. These results were motivated partly by recent experiments with [2]- and [3]catenane molecules \cite{leih-03,catenane,motor-book}. 
Such molecules are made of interlocked rings, as shown in Fig.~\ref{catenane} ([n]catenane is made of n rings.). 
By changing external conditions periodically, one can modulate the coupling strengths of smaller rings to special sites (stations) on
 the larger ring to force the smaller ones to orbit around the center of the larger ring while remaining
interlocked with it. 
Astumian  showed in \cite{astumian-07pnas}  that  adiabatic modulation of couplings to stations cannot be used to
 select a preferred mean orientation (clockwise or counterclockwise) of the orbit of the small ring in [2]catenane about the
large ring, but can be used to select a preferred mean orientation for the orbit of the pair of small rings around the large ring in a [3]catenane in Fig.~\ref{catenane}.

Any finite Markov chain can be represented as a graph. In such a graph, vertices correspond to discrete states of a system and links correspond to allowed transitions.
We say that kinetic rates $k_{ji}$ of transitions from the state $i$ to the state $j$ satisfy the {\it detailed balance condition} if they can be parametrized
by parameters $E_i$ (called {\it well depths}) and $W_{ij}=W_{ji}$ (called {\it barrier heights}) such that $k_{ji}=k\exp[E_i-W_{ij}]$. The detailed balance conditions guarantee that
if all parameters are time-independent then the state probability vector relaxes to a Boltzmann distribution. Every vertex in a graph, representing a Markov chain,
is associated with parameter $E_i$ and every link corresponds to some finite $W_{ij}$.

Rahav, Horowitz and Jarzynski \cite{jarzynski-nopump} showed that the result observed in catenane molecules is a consequence of a much more general 
{\it No-Pumping Theorem}.
They showed that for a periodic {\it non-adiabatic}
 driving protocol,  the average flux $Q$, passed through any link $i-j$ of a graph representing a finite Markov chain, 
can be written as a sum of geometric and dynamic parts, i.e. 
\begin{equation}
Q=\int_0^T dt J_{\rm dyn}(t)+\oint_{\bf c}{\bf A}\cdot d{\bf p}.
\label{no-pump1}
\end{equation}
In the first term in (\ref{no-pump1}), $T$ is the driving period.
The second term in (\ref{no-pump1}) is geometrical and depends only on the path ${\bf c}$ in the space of values of the state probability vector ${\bf p}$.
The representation (\ref{no-pump1}) is not an 
explicit solution because the evolution of the probability vector ${\bf p}$ is not assumed to be known.
 However, the authors of \cite{jarzynski-nopump} showed that when detailed balance conditions are
imposed, even on time-dependent kinetic rates, the dynamic part in (\ref{no-pump1}) becomes identically zero. Thus, even for rapid variation of the control parameters,
 the flux is purely geometrical.
 Moreover, they showed that the connection ${\bf A}$ is independent of $E_i$.
As a consequence, if some well depths $E_i$ are varied, while the $W_{ij}$ remain fixed, and if the probability vector returns to the initial values at the end of the evolution, 
the geometric term becomes also zero because $\oint_{\bf c}{\bf A}\cdot d{\bf p}={\bf A}\cdot\oint_{\bf c} d{\bf p}=0$.

\begin{figure}[h]
\centerline{\includegraphics[width=6 cm]{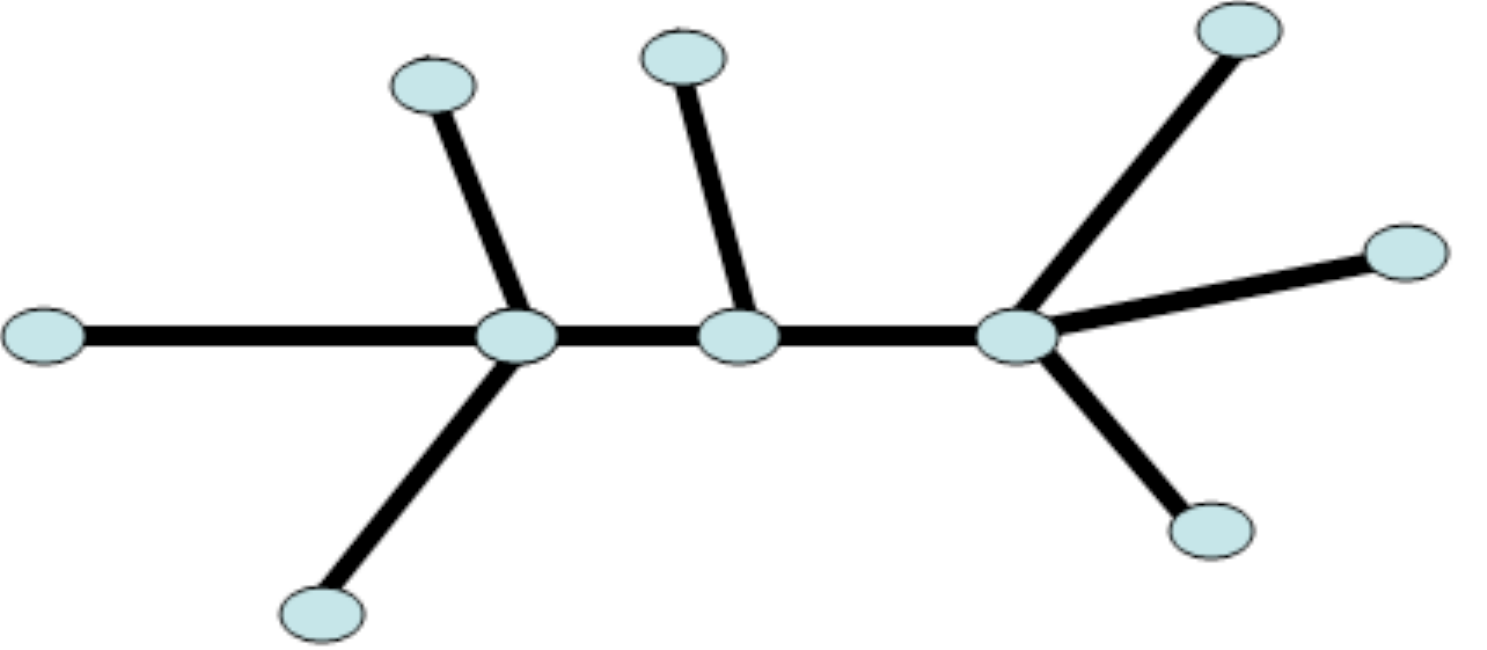}}
\caption{\label{tree} A graph with a tree-like topology.}
\end{figure}

Chernyak and Sinitsyn \cite{sinitsyn-08prl} derived and proved the Pumping-Restriction Theorem, which includes the No-Pumping Theorem of \cite{jarzynski-nopump} as a special case.
The {\it Pumping-Restriction Theorem} makes two assertions. The first assertion tells us that not all pumped currents, 
i.e. time-averaged currents induced by a cyclic evolution of parameters,   are independent. 
More precisely, pump currents through various links can be considered as vectors in a vector space whose dimension is equal to the maximum number of  time-dependent barriers $W_{ij}$ such that  
removing the links, corresponding to these barriers,
 does not break the graph into  disjoint components.  A trivial consequence of this theorem is that the pump current through any link  on a tree-like 
graph, such as the one in Fig.~\ref{tree}, is exactly zero. Indeed, removing any link from a tree (not touching the vertices) breaks the graph into disjoint components, and according to the Pumping-Restriction Theorem, the pump current on such a graph should be zero.

\begin{figure}
\centerline{\includegraphics[width=13 cm]{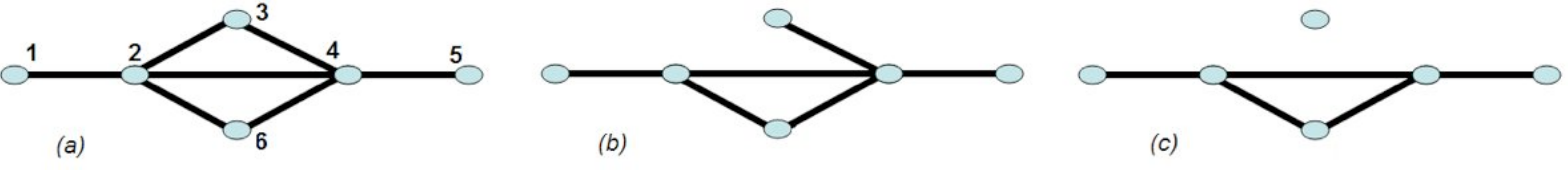}}
\caption{\label{markov6} (a) A graph representing a six-state Markov chain. (b) Removing the link $2-3$ alone does not break the graph into disjoint components. (c) Removing links $2-3$ and $3-4$ breaks the graph.}
\end{figure}

The result for a tree graph is expected because, for periodic evolution of parameters, the flux through any link on such a graph should eventually return through the same link
moving in the opposite direction since initial and final state probability vectors coincide. Hence, the integrated current must be zero. The Pumping-Restriction Theorem
leads also to many less obvious predictions. For example, in the graph in Fig.~\ref{markov6}(a), one can, by perturbing periodically  well depths $E_i$ and also only the barriers related to links $2-3$ and $3-4$,
 induce a nonzero pump current, because removing any one of the links $2-3$ or $3-4$ does not break the graph into disjoint components, as it is shown in Fig.~\ref{markov6}(b).
 According to the theorem, however, a pump current through any link in Fig.~\ref{markov6}(a) will then be
proportional to the pump current through the link $2-3$ with a constant proportionality coefficient, which is independent of the driving protocol,
 because removing both links $2-3$ and $3-4$ we would break the graph into disjoint parts, as it is shown in Fig.~\ref{markov6}(c). Hence 
the dimension of the pump current space in this case is only 1.  
The Pumping-Restriction Theorem also predicts that an arbitrary periodic driving of parameters on the links $1-2$ and $4-5$ in Fig.~\ref{markov6} alone cannot induce the pump effect, because 
removing any of those links would break the graph. 
That should suffice
to explain and illustrate the first assertion of the
Pumping-Restriction Theorem. As for the
second assertion of the theorem, it states that if pump currents are allowed, and if there are restrictions on their values predicted by the first part of the theorem,
then these  restrictions do not depend on parameters $E_i$.  

The existence of exact results, such as the No-Pumping and the 
Pumping-Restriction Theorems, tells us that there are strong constraints which must be satisfied by control parameters in order to induce a directed motion of a nanoscale device.
At this stage it is unclear whether these theorems can be extended to include 
current fluctuations, or whether the Pumping-Restriction Theorem can be related to other exact results in non-equilibrium thermodynamics, such as the Jarzynski equality \cite{jarzynski-review},
or the invariant quantities in a shear flow found in \cite{baule-08}. 
We note that a number of fluctuation theorems have been found for applications to ratchet systems and molecular motors \cite{andrieux-07,astumian-fdt,hill-book,harris-ft,astumian-pra89,astumian-08prl}, but their connections to 
geometric phases are unknown.

\section{Geometric phases and molecular motors}

Applications  of geometric phases in mathematical robotics are based on the same geometric structure as
the one discussed by Shapire and Wilczek to describe the
locomotion of living cells
 \cite{wilczek-88,shapire-88}. As in the case of  living organisms,
the internal motions of robots are confined to a
particular region as the robot
  performs its tasks. Nevertheless, as the robot interacts with its environment, periodic changes in the internal control variables can lead to 
nonperiodic effects on the robot position or on its surroundings. 
The reader is referred
to  \cite{robot-book,kelly-94,marsden-book} for an introduction to the geometric theory of
robotic motion and for references to its extensive literature.
In this section we demonstrate how geometric phases in stochastic kinetics appear in the control theory of the molecular machines.

Many biological molecules resemble motors, and
sometimes operate according to principles similar to those
which govern
 the macroscopic machines used by humans. Molecular motors are ubiquitous in living organisms.
They are employed, for example, for  transport, for injecting viruses into living cells, for unzipping the DNA, and 
for storing energy  \cite{cell,today}.

Experimental progress in the  synthesis and observation of molecular motors has been  remarkable.  
The reviews \cite{motor-chem,motor-chem2} describe many recently synthesized molecules such as rotaxanes and catenanes, which are able to perform  prescribed mechanical movements in response to external stimulus. 
Moreover, the experimental techniques for observing  molecular motion have reached a  level where  discrete steps in a molecular motor operation can be observed \cite{english-06,single-mol1}. 
It has become feasible to measure
noise  and even the tails of the probability distributions of reaction events  \cite{english-06,bezrukov,szabo-06}.

\begin{figure}[h]
\centerline{\includegraphics[height=2.4in]{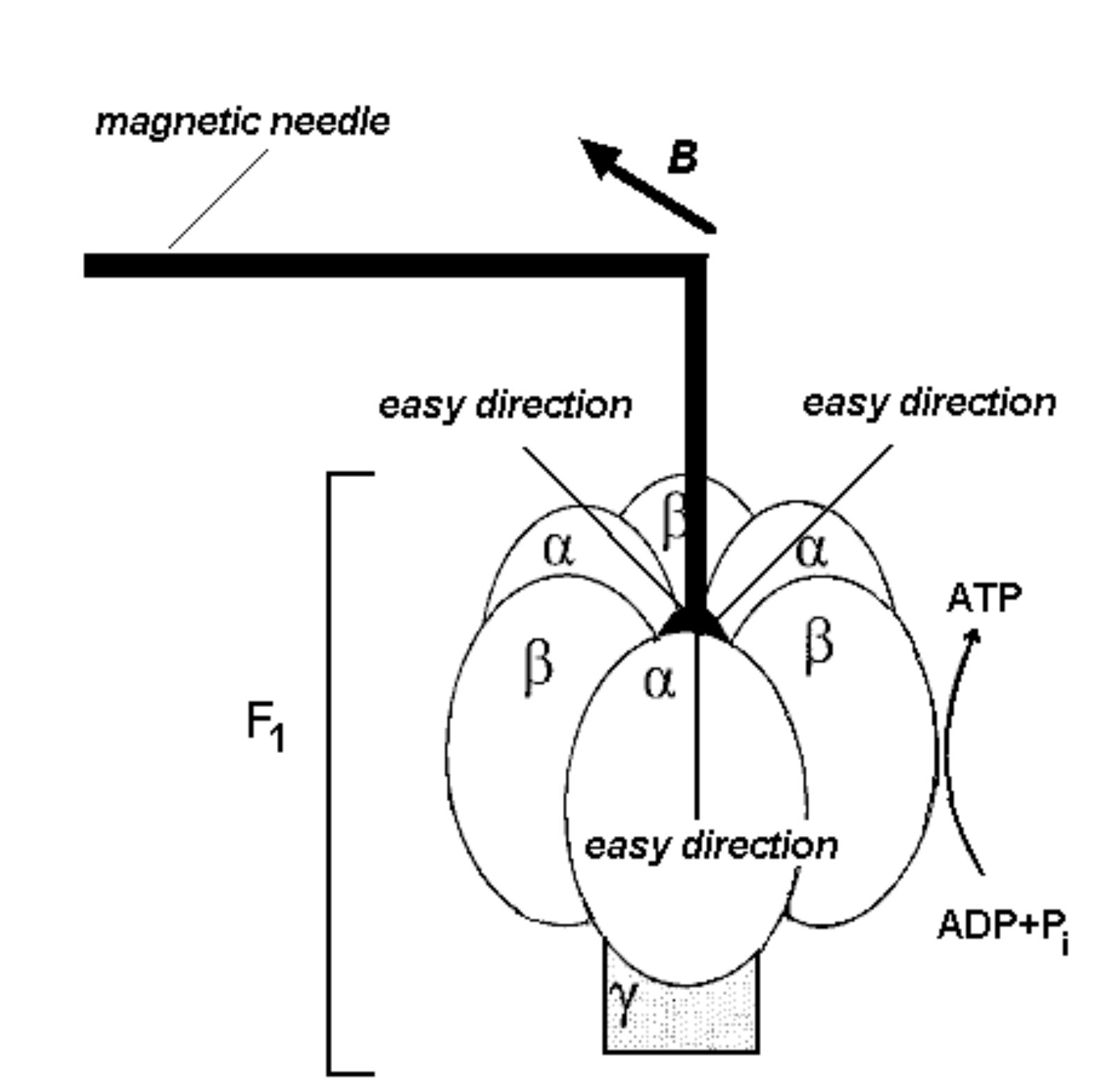}}
  \caption{\label{f0f1} F$_1$-subunit of the natural molecular motor F$_0$F$_1$-ATPase with attached magnetic needle in external magnetic field ${\bf  B}$.}
\end{figure}

Better understanding of the working principles
of nanoscale machines will make it possible
 to rebuild them  to perform
specialized tasks. 
In living cells, the molecular motor F$_0$F$_1$-ATPase can
 convert the energy of an H$^+$ gradient into chemical energy
stored in ATP.
This natural molecular motor was rebuilt recently by modifying its
F$_1$-subunit and was used to store the work done by an external
magnetic field as
chemical energy \cite{japan}.
Another artificial modification of this molecular motor, shown in Fig.~\ref{f0f1}, was used in \cite{F1-propeller} to demonstrate the rotation of a metallic bar, powered by ATP.

At the molecular level,
fluctuating forces are considerably stronger than typical external fields. Unlike macroscopic machines, molecular motors such as F$_0$F$_1$-ATPase absorb chemical energy 
 in discrete portions and randomly. 
This leads to the so called {\it shot noise} in their operations.
Molecular motors are also affected by strong thermal fluctuations. These fluctuations are not merely a theoretical
complication. For example, vesicle transport via the  ``hitchhiking'' mechanism, proposed  in \cite{nelson-08} 
depends on stochasticity in
an essential way. Single molecule experiments show that noise measurements can
provide important information about molecular structure \cite{english-06} which can, in turn, be employed to uncover details of the working cycle of a molecular motor.
Thus, the theory of molecular motor operations must take stochastic effects into account.

Much effort has been
devoted to modeling
 the thermodynamics of molecular motors using the models of stochastic pumps
 and ratchets \cite{westerhoff-86, astumian-91, astumian-90,astumian-jpa05,tsong-07}. This work has been
 reviewed in \cite{motor-1rev,reimann-02,julicher-etal-02,astumian-bio98}.
In contrast, appreciation of the role of
geometric phases in molecular motor operations is relatively recent and is still in an early stage of development
 \cite{sinitsyn-07epl,sinitsyn-07prl,parrondo-98,astumian-07pnas,magnasco-94,Qian,astumian-02pt}.

To illustrate that role,
 we consider a model inspired by the structure shown in  Fig.~\ref{f0f1}, where the metallic bar is magnetic and the whole structure is placed in a rotating magnetic field. 
The molecule has an approximate
3-fold symmetry (except for its rotating $\gamma$-subunit). If the magnetic coupling to the needle is weaker than the size of the potential barrier $W$ between any pair of metastable states,
the kinetics can be minimally described by a 3-state model with some stochastic transition rates between any pair of neighboring states, as shown in Fig.~\ref{three-state}.
Near equilibrium, rotational steps of the $\gamma$-subunit 
in both directions were observed experimentally \cite{hayashi}. 
Due to the magnetic needle, it is possible to control the relative energies of the 3 states and of the potential barriers, which separate them, by applying an external  rotating magnetic field.
We will be interested in the number of full rotations of the needle, and hence of the  $\gamma$-subunit attached to it, as it is driven in a neighborhood of thermodynamic equilibrium.

One can parametrize kinetic rates of an arbitrary Markov chain with detailed balance
conditions in the following way, using the terminology "well-depth"
and "barrier height" which was
introduced in section 9.3:
 for sites $i$ and $j$, the
transition rate from $j$ to $i$ is
$k_{ij} = k e^{E_j-W_{ij}}$, where $E_j$ is the well-depth $j$, and $W_{ij}=W_{ji}$ is the barrier height between sites $i$ and $j$. Here, the energy scale is $k_BT=1$, where
$T$ is temperature and $k_B$ is the Boltzmann constant. 
The parameter $k$ is a constant rate coefficient which sets the time scale and depends on the properties of the solution, i.e. on environment of the molecular motor.
Imitating
 the procedure that  was used  in section 5, one 
 finds that the moment generating function of the number of full rotations in the clockwise direction  (counting counterclockwise rotations with a negative sign)
 is again given by Eq. (\ref{pdf2}), but with the new Hamiltonian that reads 
\begin{equation}
\hat{H}(\chi)=  \left(\begin{array}{lll}
-(k_{21}+k_{31} ) & k_{12} e^{-i\chi/3} & k_{13}e^{i\chi/3}\\
k_{21}e^{i\chi/3}  & -(k_{32}+k_{12}) &  k_{23}e^{-i\chi/3}\\
 k_{31}e^{-i\chi/3} &  k_{32}e^{i\chi/3} & -(k_{13}+k_{23})
\end{array} \right). \quad 
\label{new_ham}
\end{equation}
The magnetic field modulates the parameters $E_i$ and $W_{ij}$, and hence $k_{ji}$.
 The 3-fold symmetry of the molecule can be taken into account by assuming the following dependence of the parameters $E_i$ on the components $B_x$ and 
$B_y$ of the magnetic field:
\begin{equation}
\begin{array}{l}
E_1=b_y,\\
E_2=b_x \cos(\pi/6) - b_y \cos(\pi/3),\\
E_3=-b_x\cos(\pi/6)-b_y \cos(\pi/3),
\end{array}
\label{et}
\end{equation}
where $b_{x/y}=-B_{x/y}|{\bf M}|$, and ${\bf M}$ is the magnetization vector of the needle. One can derive (\ref{et}) by assuming that the first metastable state corresponds
 to the magnetic bar pointing along $y$-axes  and the other two 
are generated by a rotation of the needle by angles $2\pi/3$ and $4\pi /3$. For the magnetization energy $\epsilon$, we use units where $\epsilon = - {\bf B \cdot M}$.
Suppose that the maxima of the potential barriers are shifted from the potential well minima by an angle $\phi$. Thus a reasonable assumption for the dependence of the
 barrier heights on the magnetic field would be
\begin{equation}
\begin{array}{l}
W_{12}=W +b_x \sin(\phi)+b_y\cos(\phi),\\
W_{23}=W +b_x \cos(\pi/6+\phi) - b_y \cos(\pi/3-\phi),\\
W_{13}=W- b_x\cos(\pi/6-\phi)-b_y \cos(\pi/3+\phi).
\end{array}
\label{wt}
\end{equation}


According to the procedures of \cite{sinitsyn-07epl}, which have been discussed in section 5,  the moment generating function of the number of full rotations of the needle is determined by the eigenvalue
with the  lowest real part and corresponding 
eigenvectors of (\ref{new_ham}). 
 While it is not convenient to study the exact expressions for eigenvectors of a $3\times 3$ matrix, it is not hard to derive the lowest cumulants of the rotation numbers by treating 
the counting parameter $\chi$ perturbatively, as in  \cite{summer-school-08}.
As in the example of the reversible ratchet, the principle of detailed balance imposes constraints which imply that, on average, the needle rotation becomes a purely geometric phase effect and
that the average number of needle rotations per cycle can be expressed as an integral over a surface 
 ${\bf S_c}$ whose boundary is the
contour  ${\bf c}$ in the $(b_{x},b_{y})$ parameter space, namely
\begin{equation} 
\langle n \rangle =\int \int_{{\bf S_c}} db_ydb_x F(b_x,b_y). 
\label{bcc}
\end{equation} 
The ``Berry curvature'' $F(b_x,b_y)$ 
determines the sensitivity of the system to  external driving forces. Regions with larger values of the function $F(b_x,b_y)$ correspond to parameter values at which system makes
more rotations, on average, in response to periodic parameter variations.
For a symmetric barrier configuration ($\phi=\pi/3$), the expression for $F(b_x,b_y)$ is 
particularly simple, namely
\begin{equation} 
F(b_x,b_y)=\frac{3\sqrt{3} e^{3\sqrt{3}b_x/2+3b_y}}{\left(e^{\sqrt{3}b_x/2}+e^{3b_y/2}(1+e^{\sqrt{3}b_x})\right)^3}. 
\label{Fbcc}
\end{equation} 
\begin{figure}
\centerline{\includegraphics[width=2.2in]{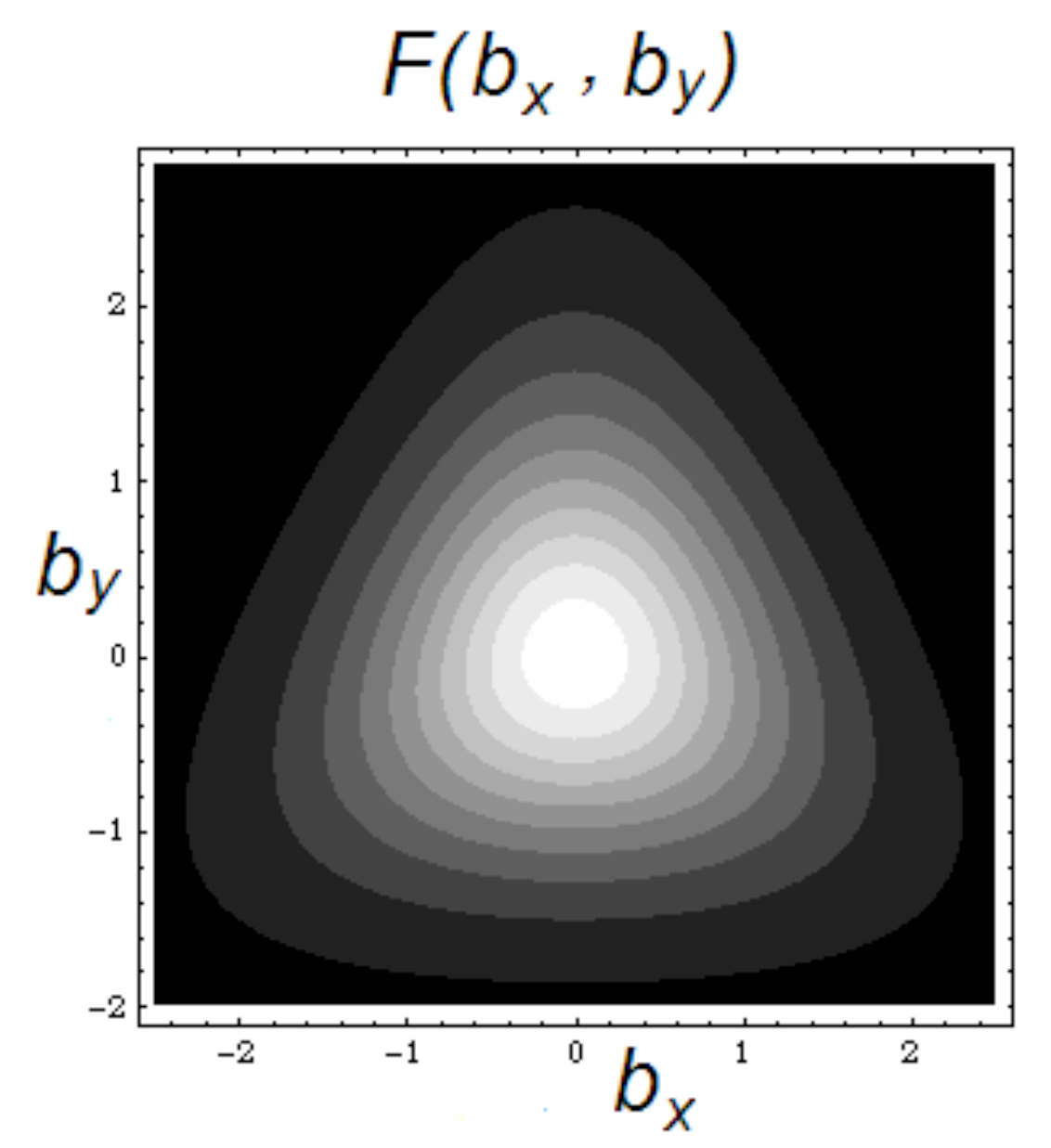}}
 \caption{\label{BerryCurv} The contour plot of the ``Berry curvature''   $F(b_x,b_y)$ from Eq. (\ref{bcc}),
as a function of control parameters $b_x$ and $b_y$. Brighter areas correspond to larger values of $F(b_x,b_y)$, and dark area corresponds to regions where  $F(b_x,b_y) \approx 0$.  }
\end{figure}
Positivity of the Berry curvature in Fig.~\ref{BerryCurv} means that the maximum number of rotations per one cycle is achieved for the contour that encloses 
the whole bright area in Fig.~\ref{BerryCurv} from a very large distance in the parameter space.
For any such a large contour, the result of the integration (\ref{bcc}) of the Berry curvature remains approximately the same and can be estimated by setting
the limits of the integration in (\ref{bcc})  to infinities, which gives
\begin{equation} 
\int_{-\infty}^{\infty} \int_{-\infty}^{\infty} db_ydb_x F(b_x,b_y)=1.
\label{FFbcc}
\end{equation} 
The result shows that 
 the system in Fig.~\ref{f0f1}  makes, on average, at most one net rotation for every rotation of the magnetic field  along a 
closed non-intersecting contour.  This result does not require that the coupling to the magnetic field be the largest energy scale in the model. In our calculations we always assumed that $W > b_{x/y} $,
 in order to use a 3-state approximation. The resulting quantization happens only on average, and the needle
is allowed to make many stochastic steps before the rotation of the magnetic field is complete.

By comparing the number of rotations of the structure with the absolute value of the field, one can determine the function $F(b_x,b_y)$  experimentally. 
 Fig.~\ref{BerryCurv} shows that $F(b_x,b_y)$ has the same 3-fold symmetry as the kinetic model in Fig.~\ref{three-state}. Its values can therefore reveal details of the internal molecular structure and
of possible components in the effective kinetic model.
Interestingly, the theory predicts that 
the geometric phase and the function in Fig.~\ref{BerryCurv} are independent of the parameter $k$ and of the size of an unperturbed barrier $W$. 
This means, in particular, that the function $F(b_x,b_y)$ can be robust against variations of the viscosity of the solution. This prediction is valid as long as the magnetic field rotation is adiabatically
slow and the system always remains close to  thermodynamic equilibrium. Thus we predict a universality of the motor response, in a sense that it does not depend on the solution viscosity,
which can be tested experimentally.
If the molecular motor is subject to additional forces, such as a proton gradient, which drive it beyond the regime of approximate thermodynamic equilibrium, this universality may no longer hold.

%
%

\section{Discussion}

The analogy between the evolution of generating functions in stochastic processes and the evolution of quantum mechanical wave functions
 allows to consider complex stochastic processes using the framework of quantum  mechanics.  
 In this review we discussed how, due to this analogy, quantum mechanical Berry phases appear to have counterparts in classical stochastic processes. 
The quantum pump effect, whose origin can be attributed to a Berry phase, has a 
stochastic counterpart with a similar geometric phase interpretation. We also showed that, as in quantum mechanics, 
geometric phases can influence the motion of coarse-grained degrees of freedom in stochastic processes after
elimination of fast variables.
This similarity raises questions about the possibility of  further analogies. 

Berry phases are responsible for a number of important effects in solid state physics, such as the quantum Hall effect. 
Although it is unclear whether or not  similar effects can be discovered in classical dissipative systems, several features of the SPE have quantum mechanical counterparts.
 We
showed in section 10 that the SPE can be quantized, i.e. the number of system rotations per cycle can be some integer. 
Similar quantization has been considered as a special feature of the quantum pump  \cite{qpump-metrology} and the quantum Hall effect \cite{niu-book}.
 There are also examples of fractionally-quantized responses of stochastic systems \cite{leih-03,astumian-07pnas}.
 Usually quantization is achieved in the limit of the maximum efficiency of the stochastic system response along
a cycle in the space of control parameters. While such limits are easy to find in simple models, little is known about how to determine them in the general case. One possibility was proposed by Shi and Niu in \cite{shi}.
Examining the diffusion  in a periodic potential they related the quantization of the stochastic ratchet current to a Chern number. It is possible that this observation can be generalized within the Olson-Ao  description of
the Bloch-Peierls-Berry dynamics \cite{Ao}. Another type of quantization was found 
in dissipative transport of a particle on a periodic lattice with non-Hermitian evolution \cite{levitov}.

Certainly, there are important differences between quantum and classical systems. The 
quantum Hall effect and the quantum theory of polarization
 require the existence of the Fermi
sea, and thus are intrinsically many-body effects which rely on the Pauli principle for a multi-particle fermionic wave function. To some extent the 
Pauli principle can be mimicked in stochastic processes by means of exclusion interactions, as
in the theory of the shot noise in electronic circuits \cite{nazarov-03}. The model we discuss in section 5, provides a simple example of a geometric phase effect which is induced by exclusion interactions.
Even in simple systems, interactions lead to important effects, such as 
violation of the assumptions of the No-Pumping Theorem \cite{jarzynski-nopump,astumian-07pnas}. 
It is therefore important  to explore geometric phases in strongly interacting many-body stochastic systems, such as in reaction-diffusion models and
multistate exclusion processes \cite{dhar-07prl,dhar-08stat}. 
In quantum field theories, Berry phases are responsible for  chiral anomalies \cite{chiral-book}. Such anomalies play an important role 
in the quantum pump effect and in the theory of the quantum Hall
effect 
 \cite{kamenev-qpump}.
 It would be 
interesting to know whether the quantum-statistical analogies discussed in this review can be used to find stochastic phenomena analogous to chiral anomalies.

  Conversely, quantum theory can benefit from the analogy with stochastic kinetics.
  For example, one can explore the possibility of a quantum mechanical analog of the Pumping-Restriction Theorem.  One can also consider geometric phases in the evolution of
the counting statistics in quantum mechanical systems \cite{golan}. Fractional quantization of pumping in stochastic systems, such as in [3]catenanes \cite{leih-03},
 may shed new light on the nature of quasiparticles in the fractional quantum Hall effect.

One of the goals of this review is to emphasize that geometric phases can play an important role in the theory of molecular motors. 
We showed that calculations of geometric phases are important in designing  molecular motors and specifying their operations. Such calculations are not based on direct numerical
solutions of differential equations with explicitly time-dependent parameters. 
Instead, by applying the theory of geometric phases it is possible to understand molecular motor operations using knowledge of only
the energy landscape of a molecule as function of control parameters. The theory also suggests new response coefficients for experimental investigation. Measurements of the Berry curvature should 
provide  new insight into the structure of motor molecules. 

Molecular machines, driven by external time-dependent forces, obey
simple universal laws, which remain to be explored and whose existence
is
indicated by the discovery of geometric phases in stochastic kinetics and exact universal results such as 
Pumping-Restriction Theorems and Fluctuation Theorems.
In the future, the science of controlled mesoscopic systems will be
transformed by the investigation of these laws.


\ack{ The author thanks Allan Adler for insightful comments, which were used to substantially improve this review. The author also thanks Qian Niu and Ilya Nemenman for useful discussions and 
Maryna Anatska for the help with illustrations.  
 This
  work was funded in part by DOE under Contract No.\ 
  DE-AC52-06NA25396.  }


\end{document}